\documentstyle[epsfig,natbib2,natbibmnfix]{mn2e}

\newcommand{\code}{{\small SPHRAY\,}}

\title[Radiative Transfer with SPHRAY]{SPHRAY: A Smoothed Particle Hydrodynamics Ray Tracer for Radiative Transfer}

\author[Gabriel Altay et al.]
      {Gabriel Altay$^{1}$, Rupert A.C. Croft$^{1}$, and Inti Pelupessy$^1$\\
       $^1$ Carnegie Mellon University, Department of Physics,
            5000 Forbes Avenue, Pittsburgh PA 15213, USA\\}

\date{Accepted 200? ???? ??.
     Received 2007 ???? ??;
     in original form 2007  xx}

\pagerange{\pageref{firstpage}--\pageref{lastpage}}
\pubyear{200?}

\newcommand{\HI}{\rm HI}
\newcommand{\HII}{\rm HII}
\newcommand{\HeI}{\rm HeI}
\newcommand{\HeII}{\rm HeII}
\newcommand{\HeIII}{\rm HeIII}
\newcommand{\xHI}{x_{\rm HI}}
\newcommand{\xHII}{x_{\rm HII}}
\newcommand{\xHeI}{x_{\rm HeI}}
\newcommand{\xHeII}{x_{\rm HeII}}
\newcommand{\xHeIII}{x_{\rm HeIII}}

\newcommand{\yHII}{y_{\rm HII}}

\newcommand{\yHeII}{y_{\rm HeII}}
\newcommand{\yHeIII}{y_{\rm HeIII}}
\newcommand{\ye}{y_{\rm e}}
\newcommand{\nHI}{n_{\rm HI}}
\newcommand{\nHII}{n_{\rm HII}}
\newcommand{\nHeI}{n_{\rm HeI}}
\newcommand{\nHeII}{n_{\rm HeII}}
\newcommand{\nHeIII}{n_{\rm HeIII}}
\newcommand{\nel}{n_{\rm e}}
\newcommand{\nH}{n_{\rm H}}
\newcommand{\nHe}{n_{\rm He}}

\newcommand{\figcapt}{Shown are the results from \code (solid red line), 
{\small CRASH} (dash-dot-dot-dot purple line), 
{\small RSPH} (dashed green line), 
{\small C$^2$-Ray} (small dashed light blue line) 
and the results from all the other codes in the comparison project that completed this test (dotted orange lines). }

\newcommand{\testone}{Comparison Project Test 1 (HII region expansion in a gas at constant density and temperature). }
\newcommand{\testtwo}{Comparison Project Test 2 (HII region expansion in a gas at constant density with varying temperature). }
\newcommand{\testthree}{Comparison Project Test 3 (I-front trapping in a dense clump). }
\newcommand{\testfour}{Comparison Project Test 4 (Multiple sources in a cosmological density field). }

\begin{document}

\maketitle

\label{firstpage}

\begin{abstract}
We introduce the publically available code \code, a Smoothed Particle
Hydrodynamics (SPH) ray tracer designed to solve the 3D, time dependent,
radiative transfer (RT) equation for cosmological density fields.  The SPH
nature of \code makes the incorporation of separate hydrodynamics
and gravity solvers very natural.
 \code relies on a Monte Carlo (MC) ray tracing scheme
that does not interpolate the SPH particles onto a grid but instead integrates
directly through the SPH kernels.  Given an arbitrary (series of) SPH density
field(s) and a description of the sources of ionizing radiation, the code will
calculate the non-equilibrium ionization and temperature state of Hydrogen
($\HI, \HII$) and Helium ($\HeI, \HeII, \HeIII$).  The sources of radiation
can include point like objects, diffuse recombination radiation, and a
background field from outside the computational volume.  The MC ray tracing
implementation allows for the quick introduction of new physics and is
parallelization friendly.  A quick Axis Aligned Bounding Box (AABB) test taken
from computer graphics applications allows for the acceleration of the
raytracing component.  We present the algorithms used in \code and verify the
code by performing the test problems detailed in the recent Radiative Transfer
Comparison Project of Iliev et. al. The source code for \code and example SPH density fields are made
available on a companion website (www.sphray.org).

\end{abstract}


\section{Introduction}
In numerical cosmology, prescriptions for the treatment of gravity and
hydrodynamics are well developed and have been validated against one another in
several comparison studies 
\citep[see][]{1999ApJ...525..554F, 2005ApJS..160....1O,
2005ApJS..160...28H, 2007arXiv0706.1270H, 2007MNRAS.374..196R,
2007MNRAS.380..963A, 2007arXiv0709.2772P}.  
The density and temperature fields they produce provide input for sub-resolution models of star formation and
feedback via supernovae \citep[e.g.][]{2003MNRAS.339..289S} and black holes
\citep[e.g.][]{2007arXiv0705.2269D}.  Numerical radiative transfer (RT)
techniques, necessary to calculate the interaction of the ionizing photons
produced by these sources with the cosmological gas, have not yet reached the
level of maturity attained by $N$-body and gas dynamics solvers.  Flexible and
accurate RT techniques, validated against analytic solutions and in comparison projects,
are necessary to properly interpret many observations and guide the
development of theoretical models from cosmological through 
stellar scales.  This is especially true
for analysis of upcoming 21 cm surveys such as 
21CMA \footnote{ http://21cma.bao.ac.cn/index.php } (formerly PAST),
LOFAR \footnote{ www.lofar.org }, 
MWA \footnote{ www.haystack.mit.edu/ast/arrays/mwa },
SKA \footnote{ www.skatelescope.org };
modeling absorption lines in the spectra of high redshift quasars and gamma ray burst afterglows,
and understanding the feedback processes which influence star and galaxy formation.

The introduction of 3D radiative transfer into cosmological simulations is
complicated by several issues.  The specific intensity $ I_{\nu} = I (
\vec{\mathbf x} , \hat{\mathbf n}, \nu, t ) $ is a function of seven variables
leading to a solution space with high dimensionality.  R-type ionization
fronts can travel at nearly the speed of light through underdense regions and
many times the speed of sound in dense regions leading to radiative time
scales orders of magnitude smaller than dynamical time scales.  In addition,
radiative transfer and hydrodynamic processes are coupled.  For example, photo
heating creates large pressure gradients near luminous sources and can modify
star formation rates while hydrodynamic temperature changes affect the
recombination, collisional ionization, and radiative cooling rates of the
cosmological gas.

In designing a numerical radiative transfer scheme, it is practical to utilize
the hydrodynamic frameworks that have already been developed.  These can
generally be categorized as Lagrangian, particle based methods \cite[see][for
a review of Smoothed Particle Hydrodynamics]{1992ARA} or Eulerian, grid based
methods \cite[see e.g.][for information on Adaptive Mesh Refinement
codes]{2004astro.ph..2230N}.  In this paper, we describe \code, a code that
performs radiative transfer calculations on Smoothed Particle Hydrodynamics
(SPH) density fields.  A flexible, general purpose, radiative transfer method
tightly coupled to SPH hydrodynamic simulations could be adapted to handle
many different astrophysical problems.  Currently, \code works on static
density fields. However, it calculates quantities that are equivalent to the
change in specific energy (or entropy) for individual SPH particles due to
photoionization/photoheating. This makes the coupling of 
\code with gravity and hydrodynamics relatively straightforward.
We leave this to future work.

SPH \citep{1977AJ.....82.1013L,1977MNRAS.181..375G} is a gridless, Lagrangian
method that discretizes a fluid into particles.  The self-gravity of
these particles can be treated in the same way as $N$-Body particles,
but they are also
subject to hydrodynamic forces.  The combination of SPH with tree
structures \citep{1989ApJS...70..419H} and of tree structures with the 
particle-mesh method \citep{1995ApJS...98..355X,2002JApA...23..185B}, provides
a very flexible computational tool.

The earliest combinations of SPH and RT were by \cite{1977AJ.....82.1013L} -
one of the papers that introduced SPH - and \cite{1985PASAu...6..207B}. These
authors modeled radiation transport as a diffusion process.  The study of
dense astrophysical regions such as molecular clouds and collapsing protostars
has continued along this line in the work of
\cite{2004MNRAS.353.1078W,2005MNRAS.364.1367W,2006ApJ...639..559V} and
\cite{2007ApJ...661L..77M}.

The highly variable optical depths through 
voids and Lyman Limit systems in
cosmological volumes do not lend themselves to a diffusion description of
radiation.  Here, either raytracing schemes or moment methods must be used.
\code uses Monte Carlo ray tracing to accomplish RT.  The simplicity of this
approach allows new physics to be included in \code very easily and provides a
framework that is conducive to parallelization.  The accuracy of raytracing
methods in general is determined by their ability to cover the simulation
volume with a sufficient number of rays.  This is a computationally 
expensive process and various approaches have been presented in the literature.

\cite{2003MNRAS.343..900O} combined a raytracing scheme with SPH in which the
emitters and absorbers of radiation are at
the deepest levels of a Barnes \& Hut
tree.  In this scheme, the internal energies of the SPH particles are
 modified by
interpolating the changes in energy of the tree leaves onto the particles and
vice versa.  These radiative calculations are made in between hydrodynamic
time steps and allow for the coupling of the two processes.
\cite{2005AA.439..153} use a similar Barnes \& Hut ray tracing scheme, but
supplement the radiative transfer cells from the tree with a further
star grid around sources.

Both of these methods, in effect, propagate photon packets through a randomly
chosen optical depth and then allow them to be absorbed or scatter. 
In this way, the
temperature and emergent spectra can be calculated, however neither method
solves for the ionization fraction explicitly but instead uses either a
temperature-opacity or a total density-opacity relation.  In addition, they
raytrace through the adaptive grids generated from the Barnes \& Hut trees and
not the SPH particles themselves.

\cite{2000MNRAS.315..713K}, taking advantage of the neighbor lists already in
place in SPH simulations, introduced a fast method to find the optical depth
from a source to a target particle.  Variations of this method have since been
utilized in \cite{2006PASJ...58..445S} and \cite{2007MNRAS.382.1759D} to
construct radiative transfer schemes in which each particle influenced by the
radiation of a source, becomes a target of that source during the radiative
update.  A decision concerning how to treat the target particle is made 
based on
the optical depth along the ray connecting the source and the
target. \cite{2007ApJ...663..687Y} have presented a related method in which
photon arrival times are calculated for each particle surrounding a given
source by integrating the ionization front jump conditions along rays.  After
the photon arrival time for a source-particle pair, the photoionization rate
is computed in the optically thin limit.  \cite{2008arXiv0802.1715P} have
recently introduced an SPH / photon packet based radiative transfer scheme in
which packets are emitted into cones around the sources and propogated among
the particles using the neighbor search lists.   

\code differs from the above methods in that it first uses a fast Axis Aligned
Bounding Box (AABB) test to find the intersections of a ray and a Barnes \&
Hut tree that stores the particles.  Next, the particles in these tree leaves
are tested to see if they intersect the ray, yielding their impact parameter.
In this way, every particle whose smoothing volume is intersected by the ray
is stored in a raylist.  In the course of moving along the ray from the
source, the ionization and temperature state of each particle is updated
leading to more particle updates per ray. In this sense, \code shares the same
ray-update ideas as the Monte Carlo ray tracing code {\small CRASH} by
\cite{2003MNRAS.345..379M}, but is applied to SPH density fields.  With a
sufficient number of rays, the native SPH resolution can be preserved.  

The format of this paper is as follows.  In \S2 we review the basic equations
governing radiative transport and the evolution of the ionization and
temperature state of a cosmological gas.  We also derive approximate analytic
solutions necessary for an iterative numerical solution.  In \S3 we describe
the algorithms used by \code.  In \S4 we present the results of several
standard RT tests.  These tests were chosen to be the same as those in a
recent radiative transfer comparison project
\citep{2006MNRAS.371.1057I}.  They include: (1) isothermal HII region
expansion; (2) HII region expansion with evolving temperature; (3) I-front
trapping and shadowing by a dense clump; (4) multiple sources in a
cosmological density field.  We make our closest
comparisons of \code with {\small
CRASH} (the only other Monte Carlo code in the project), and a code 
described in \cite{2006PASJ...58..445S} called {\small RSPH} (the
only other SPH code in the project).  Finally, in
\S5 we discuss future applications and improvements of \code as well as
summarizing and discussing our results.

\section{Basic Physics - Radiative Transfer, Ionization, and Temperature Evolution}

In this section
we review the equations of 3D radiative transfer, the ionization and
temperature equations for a cosmological gas, and the approximations that are
made in \code.  To facilitate comparison with other radiative transfer
codes we review some of the other approximations which can be made.

\subsection{Notation}

In what follows, we use Roman numerals to indicate the ionization state of an
element (H,He) in the standard way.  Elements without Roman numerals refer to
the nuclei of atoms (or all ionization states).  A subscripted $n$ refers to
the number density of an element (or a specific ionization state of an
element).  A subscripted $x$ refers to the ratio of the number density of a
specific ionization state to the number density of all nuclei of that element.
A subscripted $y$ refers to the ratio of the number density of the subscripted
species to the number density of H nuclei, for example,

\begin{eqnarray}
\xHeII &=& \frac{\nHeII}{\nHe} \\
\yHeIII &=& \frac{\nHeIII}{\nH}
\end{eqnarray}

\subsection{Radiative Transfer Equation}


The 3-D radiative transfer equation in a frame comoving with the expansion of
the Universe can be written \citep[e.g.][]{1998MmSAI..69..455N},

\begin{equation}
\frac{1}{c} \frac{\partial I_{\nu}}{\partial t} + \frac{ {\mathbf {\hat{n}}} \cdot \nabla I_{\nu}}{\bar{a}} - 
\frac{H}{c} \left( \nu \frac{\partial I_{\nu}}{\partial \nu} - 3 I_{\nu} \right) = 
\epsilon_{\nu} - \kappa_{\nu} I_{\nu}
\end{equation}

where $ \epsilon_{\nu} $ and $ \kappa_{\nu} $ are the emission and extinction
coefficients respectively, $H = \dot{a}/a$ is the Hubble parameter, $\bar{a} =
a / a_e $ is the scale factor at time $t$ divided by the scale factor
at time $t_e$ (when the photons in the ray were emitted), and $I_{\nu} = I
({\mathbf {\vec{x}}}, {\mathbf {\hat{n}}}, \nu, t) $ is the specific
intensity.

For photons with a mean free path $\lambda_{\rm mfp}$ much less than the
Horizon size $c/H$, the classical radiative transfer equation is a valid
approximation.

\begin{equation}
\frac{1}{c} \frac{\partial I_{\nu}}{\partial t} + \hat{\mathbf n}  \cdot \nabla I_{\nu} = 
\epsilon_{\nu} - \kappa_{\nu} I_{\nu}
\end{equation}

This local approximation holds fairly well before the percolation stage of reionization
when the growing ionization bubbles are still insulated from each other by the
optically thick IGM.  Care must be taken once the majority of the IGM is
reionized and becomes optically thin allowing photons to travel distances greater than
the simulation box length.  The effect of these background fluxes from
outside the simulation volume must be taken into account, especially for high energy photons which have
longer mean free paths and the potential to ionize and heat the IGM
after being redshifted.  The treatment of these non-local fluxes should be
tailored to the specific problem at hand and so were not 'hard-wired' into
\code.  For the test cases presented in \S 4 they were not necessary.

Another caveat to using the classical equation, as explained in
\cite{1999ApJ...523...66A}, is that it is only valid when $ |\nu \partial
I_{\nu} / \partial \nu| \leq I_{\nu} $ and hence only for continuum radiation.
However, the classical equation can still be used for line radiation if the
redshifted absorption (photo-ionization) cross-sections are used when
determining $\kappa_{\nu}$.

If $\epsilon_{\nu}$ and $\kappa_{\nu}$ can be approximated as constant, a time independent RT equation can be used.

\begin{equation}
\hat{\mathbf n} \cdot \nabla I_{\nu} = \epsilon_{\nu} - \kappa_{\nu} I_{\nu}
\end{equation}

This is a good approximation for individual SPH particles over a sufficiently
short time, however \citep[as is also discussed in][]{1999ApJ...523...66A} it
breaks down close to sources and allows the possibility of ionization
fronts that travel faster than the speed of light.  This can be quantified by
examining the ionization front jump condition for a single point source
ionizing a uniform density, constant temperature, Hydrogen gas,

\begin{equation} 
\nH \frac{dr_{I}}{dt} = \frac{\dot{N}}{4 \pi r_{I}^2} - \alpha_{\rm H} \int_0^{r_{I}} \nel \nH \xHII dr
\end{equation}

where, $r_{I}$ is the distance to the ionization front from the source,
$\dot{N}$ is the number of photons per second emitted by the source, and
$\alpha_{\rm H}$ is the recombination rate.  An upper limit on the radius,
$r_c$ within which the ionization front has a speed greater than c is,

\begin{equation}
r_c \leq \sqrt{\frac{\dot{N}}{4 \pi \nH c} }. 
\end{equation}

Within this region, use of the time independent equation breaks down.  In a
raytracing scheme, this can be avoided by stopping rays once they have reached
a distance $d = ct_{on}$ where $t_{on}$ is the amount of time the source has
been on.  The photons that were in the ray can be saved and traced from the
stopping point once enough time has elapsed.  In practice this is not always
necessary.  For example, the first test presented in \S 4 has $r_c/r_s = 6.9
\times 10^{-3}$ where
 the Str{\"o}mgren radius, $r_s = 5.4$ kpc, 

In \code, the diffuse component of the radiation field is modeled using the on-the-spot (OTS) approximation, or as a set
of many point sources and so for all calculations we can set
$\epsilon_{\nu}=0$ along the ray, further simplifying the RT equation,

\begin{equation}
\frac{\partial I_{\nu}}{\partial r} =  - \kappa_{\nu} I_{\nu}
\end{equation}

which has the analytic solution, 
\begin{equation} 
I_{\nu}(r) = I_{\nu}(r_0) e^{-\tau(r)} 
\end{equation}

where

\begin{equation}
\tau = 
\int_0^r n_x(r) \, \sigma(\nu) \, dr = 
\int_0^r \kappa_{\nu} \, dr
\end{equation}

In principle, $\kappa_{\nu}$ should include contributions from every process
that removes photons from the ray under consideration (photo absorption,
Thomson scattering, dust, etc.).  For the tests presented here, we consider
only photo absorption, however it would be straightforward to add terms to
account for other processes.

\subsection{Ionization Equations}
In this section we review the equations that determine the time development of
the ionization fractions.  They represent the contributions from
photo-ionization, collisional ionization and recombination.  Analytic and time
averaged solutions in the case of constant rates are derived for use in an
iterative solution scheme which relaxes the stringent constraints on the time
step.

\subsubsection{Chemistry}

\code follows the non-equilibrium evolution of six species [$ \xHI, \xHII,
\xHeI, \xHeII, \xHeIII, \ye $], only three of which are independent.

\begin{equation}
\xHI + \xHII = 1
\end{equation}

\begin{equation}
\xHeI + \xHeII + \xHeIII = 1
\end{equation}

\begin{equation}
\ye = \yHII + \yHeII + 2 \yHeIII + y_{\rm Z}
\end{equation}

where $ y_{\rm Z} $ represents a constant background of free electrons from
ionized metals.  This background has a negligible effect on the evolution of
the ionization fractions of H and He, but provides stability in the case of
very small levels of ionization.  We note that this is a subset of all atomic
species relevant to primordial chemistry.  Although inclusion of species
involved in the formation of molecular Hydrogen [ {\small H$^-$, H$_2$,
H$_2^+$}] is important in studies of primordial star formation, they have only
a small impact on the evolution of the IGM.  Primordial gas chemistry is discussed in detail in \cite{1997NewA....2..209A} and \cite{1997NewA....2..181A}.

\subsubsection{Differential Equations}

The processes that we will consider in the evolution of the ionization
fractions are, recombination $\alpha_I$, collisional ionization $\gamma_A$,
and photo-ionization $\Gamma_A$ where $ A \in \{$ {\small HI, HeI, HeII} $\} $
is one of the photo absorbing species and $ I \in \{$ {\small HII, HeII,
HeIII} $\} $ is one of the photo-ionized species.  The equations can be
written down directly,

\begin{eqnarray}
\frac{d\xHI}{dt} &=& - G_{\rm HI} \xHI + R_{\rm HII} \xHII  \\ 
\frac{d\xHII}{dt} &=& G_{\rm HI} \xHI - R_{\rm HII} \xHII  
\end{eqnarray}

\begin{eqnarray}
\frac{d\xHeI}{dt} &=& - G_{\rm HeI} \xHeI + R_{\rm HeII} \xHeII \\  
\frac{d\xHeII}{dt} &=& G_{\rm HeI} \xHeI - (G_{\rm HeII} + R_{\rm HeII}) \xHeII + \nonumber \\
      && R_{\rm HeIII} \xHeIII  \\ 
\frac{d\xHeIII}{dt} &=&  G_{\rm HeII} \xHeII - R_{\rm HeIII} \xHeIII 
\end{eqnarray} 

where we have grouped the ionizing terms (
writing $G_{A} = \Gamma_{A} + \gamma_{A}
\nel $) together, and included the electron number density in the
recombination term ($R_{I} = \alpha_{I} \nel $).  In matrix form,

\begin{eqnarray}
\dot{\mathbf x}_{\rm H}  &=& {\mathbf M}_{\rm H} {\mathbf x}_{\rm H} \\
\dot{\mathbf x}_{\rm He} &=& {\mathbf M}_{\rm He} {\mathbf x}_{\rm He}
\end{eqnarray}

where, 

\begin{equation}
{\mathbf M}_{\rm H} = \left(
\begin{array}{cc}
 - G_{\rm HI} &  R_{\rm HII} \\
   G_{\rm HI} & -R_{\rm HII}
\end{array} \right) 
\end{equation}

\begin{equation}
{\mathbf M}_{\rm He} = \left( 
\begin{array}{ccc} 
 - G_{\rm HeI} &  R_{\rm HeII} & 0 \\
   G_{\rm HeI} & -( G_{\rm HeII} + R_{\rm HeII} ) & R_{\rm HeIII} \\
  0                   & G_{\rm HeII}  &  -R_{\rm HeIII}
\end{array} \right)
\end{equation}

In general, every species with $N_{\rm is}$ ionization states leads to an $N_{\rm is} \times N_{\rm is}$ tridiagonal matrix.

\subsubsection{Analytic Solutions}

In order to proceed with a straightforward numerical integration, the
equations in the previous section are sufficient.  However, the time steps are
restricted by the stiff nature of the differential equations and so \code can
also be run with an iterative ionization solver.  This solver is based on the
method used in the code {\small C$^2$-Ray} 
presented by \cite{2006NewA...11..374M}
where the detailed Hydrogen solution is given.  For completeness we give the
Helium solutions as well.  The specific implementation in \code is described
in \S3.  It requires time averaged analytic solutions for the ionization
fractions which are derived below.

If we assume that the  $G_{A}$ and $R_{I}$ are constant the analytic solutions have the following form, 

\begin{eqnarray}
\xHI(t) &=& \xHI^{eq} + C_{\rm H}^1 e^{\nu t} \\
\xHII(t) &=& \xHII^{eq} + C_{\rm H}^2 e^{\nu t}
\end{eqnarray}

\begin{eqnarray}
\xHeI(t)   &=& \xHeI^{eq}   + C_{\rm He}^1 e^{\lambda_1 t} + C_{\rm He}^2 e^{\lambda_2 t}    \\
\xHeII(t)  &=& \xHeII^{eq}  + C_{\rm He}^3 e^{\lambda_1 t} + C_{\rm He}^4 e^{\lambda_2 t}   \\
\xHeIII(t) &=& \xHeIII^{eq} + C_{\rm He}^5 e^{\lambda_1 t} + C_{\rm He}^6 e^{\lambda_2 t}
\end{eqnarray}

with the equilibrium solutions given by,

\begin{eqnarray}
\xHI^{eq}  &=& \frac {R_{\rm HII}} { G_{\rm HI} + R_{\rm HII} } \\
\xHII^{eq} &=& \frac {G_{\rm HI }} { G_{\rm HI} + R_{\rm HII} } 
\end{eqnarray}

\begin{eqnarray}
\xHeI^{eq} &=& \frac {R_{\rm HeII} R_{\rm HeIII}} 
{R_{\rm HeII} R_{\rm HeIII} + R_{\rm HeIII} G_{\rm HeI} + G_{\rm HeI} G_{\rm HeII}}  \\
\xHeII^{eq} &=& \frac {R_{\rm HeIII} G_{\rm HeI}} 
{R_{\rm HeII} R_{\rm HeIII} + R_{\rm HeIII} G_{\rm HeI} + G_{\rm HeI} G_{\rm HeII}}  \\
\xHeIII^{eq} &=& \frac {G_{\rm HeI} G_{\rm HeII}} 
{R_{\rm HeII} R_{\rm HeIII} + R_{\rm HeIII} G_{\rm HeI} + G_{\rm HeI} G_{\rm HeII}}  
\end{eqnarray}

Each system has one eigenvalue equal to zero corresponding to the equilibrium
solutions.  The non-zero eigenvalues and the other constants can be expressed
in terms of the $G_{A}$ and $R_{I}$.  For Hydrogen,

\begin{eqnarray}
C_{\rm H}^1 &=& \Delta \xHI  = \xHI^0  - \xHI^{eq}  \\
C_{\rm H}^2 &=& \Delta \xHII = \xHII^0 - \xHII^{eq} \\
\nu &=& -(G_{\rm HI} + R_{\rm HII}) \\
\end{eqnarray}

where $\xHI^0$ and $\xHII^0$ are initial values and $\nu$ is the non-zero
eigenvalue for the Hydrogen system.  For Helium, the expressions contain more
terms, but they can be easily written down with the use of some notation.  The
matrix ${\mathbf S}_{\rm He}$ of eigenvectors that will diagonalize $ {\mathbf
M}_{\rm He}$ is,

\begin{eqnarray}  
{\mathbf S}_{\rm He}  = 
{\mathbf S}_{\rm He}^{i,j} =
\left(
\begin{array}{ccc}
\frac{ R_{\rm HeII}}{G_{\rm HeI}} & \frac{ R_{\rm HeII}}{\lambda_1 + G_{\rm HeI}} & \frac{ R_{\rm HeII}}{\lambda_2 + G_{\rm HeI}} \\
1 & 1 & 1 \\
\frac{G_{\rm HeII}}{R_{\rm HeIII}} & \frac{G_{\rm HeII}}{\lambda_1 + R_{\rm HeIII}} & \frac{G_{\rm HeII}}{\lambda_2 + R_{\rm HeIII}} 
\end{array} \right)
\end{eqnarray}

where $\lambda_1$ and $\lambda_2$ are the non-zero eigenvalues of the Helium system.  

\begin{eqnarray}
\lambda_1 &=& -(s+p) \\
\lambda_2 &=& -(s-p)
\end{eqnarray}

with

\begin{eqnarray}
s &=& \frac{1}{2} (R_{\rm HeII} + R_{\rm HeIII} + G_{\rm HeI} + G_{\rm HeII}) \\
d &=& R_{\rm HeII} R_{\rm HeIII} + R_{\rm HeIII} G_{\rm HeI} + G_{\rm HeI} G_{\rm HeII} \\
p &=& \sqrt{s^2-d} 
\end{eqnarray}

We require the six constants $C_{\rm He}^i$ in terms of the values $G_{A}$ and
$R_{I}$. Because the Helium system is constrained by the three differential
equations and the fact that the ionization fractions must sum to one, there is
some choice in the way we do this.  \code follows $\xHeII$ and $\xHeIII$ and
so a convenient relation is,

\begin{eqnarray}
C_{\rm He}^1 = b {\mathbf S}_{\rm He}^{1,2} &
\quad  C_{\rm He}^3 = b \quad &  
C_{\rm He}^5 = b {\mathbf S}_{\rm He}^{3,2} \\
C_{\rm He}^2 = c {\mathbf S}_{\rm He}^{1,3} & 
\quad  C_{\rm He}^4 = c  \quad & 
C_{\rm He}^6 = c {\mathbf S}_{\rm He}^{3,3}  
\end{eqnarray}

with, 

\begin{eqnarray}
b &=& \frac{\Delta \xHeIII -  \Delta \xHeII {\mathbf S}_{\rm He}^{3,3} }
{ {\mathbf S}_{\rm He}^{3,2} - {\mathbf S}_{\rm He}^{3,3} }  \\
c &=& \frac{\Delta \xHeII {\mathbf S}_{\rm He}^{3,2} - \Delta \xHeIII  }
{ {\mathbf S}_{\rm He}^{3,2} - {\mathbf S}_{\rm He}^{3,3} }
\end{eqnarray}

At the heart of this iterative method is the use of time averaged ionization
fractions, optical depths, and photo-ionization rates to take larger time
steps than would normally be possible.  The time averaged ionization fractions
$\langle x \rangle = \frac{1}{\Delta t}\int_0^{\Delta t} x (t) dt $ of the
above solutions are,

\begin{eqnarray}
\langle \xHI \rangle &=& \xHI^{eq} + \frac{C_{\rm H}^1}{\nu} \left( e^{\nu \Delta t}-1 \right) \frac{1} {\Delta t} \\ 
\langle \xHII \rangle &=& \xHII^{eq} + \frac{C_{\rm H}^2}{\nu} \left( e^{\nu \Delta t}-1 \right) \frac{1} {\Delta t}.
\end{eqnarray}

\noindent $ \langle \xHeI \rangle = \xHeI^{eq} + $
\begin{equation}
\quad \left[ 
\frac{C_{\rm He}^1 \left( e^{\lambda_1 \Delta t} - 1 \right)}{\lambda_1 \Delta t} +
\frac{C_{\rm He}^2 \left( e^{\lambda_2 \Delta t} - 1 \right)}{\lambda_2 \Delta t} 
\right] 
\end{equation}

\noindent $ \langle \xHeII \rangle = \xHeII^{eq} + $
\begin{equation}
\quad \left[ 
\frac{C_{\rm He}^3 \left( e^{\lambda_1 \Delta t} - 1 \right)}{\lambda_1 \Delta t} +
\frac{C_{\rm He}^4 \left( e^{\lambda_2 \Delta t} - 1 \right)}{\lambda_2 \Delta t} 
\right] 
\end{equation} 

\noindent $ \langle \xHeIII \rangle = \xHeIII^{eq} + $
\begin{equation}
\quad \left[
\frac{C_{\rm He}^5 \left( e^{\lambda_1 \Delta t} - 1 \right)}{\lambda_1 \Delta t}  + 
\frac{C_{\rm He}^6 \left( e^{\lambda_2 \Delta t} - 1 \right)}{\lambda_2 \Delta t} 
\right] 
\end{equation}

\subsection{Temperature Equation}

There are three terms in the temperature evolution equation.  One for the
photo heating ${\mathcal H}$, one for various atomic cooling \footnote{note
that Compton scattering can have a negative contribution to the cooling
function if the temperature of the background radiation field is greater than
the gas kinetic temperature} processes $ \Lambda $, and one for the change in
temperature due to the change in the number of free particles.

\begin{equation}
\frac{ d T }{ d t } = \frac {2}{3  n  k_B} ( {\mathcal H} - \Lambda ) - \frac{T}{n} \frac{dn}{dt}
\end{equation}
 
\subsubsection{Photo Heating Term}

The term
${ \mathcal H} $ accounts for the kinetic energy of the photoionized electrons
which quickly gets transferred to the other particle species.  
Let us suppose that a unit
volume of gas absorbs $\dot{N}_{\gamma}$ photons per second.  The fraction of
absorption due to a given particle species is proportional to the optical
depth of that species through the volume.  The photo-heating rate for a
monochromatic ray, ${\mathcal H} = {\mathcal H}_{\rm HI} + {\mathcal H}_{\rm
HeI} + {\mathcal H}_{\rm HeII}$ can be simply expressed in this way,

\begin{equation} 
{\mathcal H}_{\rm HI} = \dot{N}_{\gamma} \frac{\tau_{\rm HI}}{\tau_{\rm all}} (h\nu - h\nu_{\rm HI})
\end{equation}

\begin{equation} 
{\mathcal H}_{\rm HeI} = \dot{N}_{\gamma} \frac{\tau_{\rm HeI}}{\tau_{\rm all}} (h\nu - h\nu_{\rm HeI})
\end{equation}

\begin{equation} 
{\mathcal H}_{\rm HeII} = \dot{N}_{\gamma} \frac{\tau_{\rm HeII}}{\tau_{\rm all}} (h\nu - h\nu_{\rm HeII})
\end{equation}

where $\nu_{\rm HI}, \nu_{\rm HeI}$, and $ \nu_{\rm HeII} $ are the ionization threshold frequencies.

\subsubsection{Atomic Cooling Term}

The atomic cooling function, $\Lambda$, includes the following physical processes, 

\begin{itemize}
\item Collisional Ionization Cooling ( $\zeta$ )
\item Collisional Excitation Cooling ( $\psi$ )
\item Recombination Cooling ( $\eta$ )
\item Bremsstrahlung Cooling ( $\beta$ )
\item Compton Heating/Cooling ( $\chi$ )
\end{itemize}

The numerical value of $\Lambda = \Lambda( n_A, n_I, \nel, T, T_{\rm bkgnd}) $ is calculated using the rates detailed in Appendix A.

\section{Numerical Techniques}

In this section we outline the numerical techniques used to solve for the
ionization and temperature state.  \code is a Monte Carlo code and is based on
sampling the radiation field along 1-d characteristics.  This is accomplished
by tracing rays that extend a predefined length.  The length can be chosen in
a number of ways.  For vacuum boundary conditions the obvious choice is to
terminate the rays at the edge of the volume.  For reflective or periodic
boundary conditions a criterion can be applied to the properties of
 photons in the ray  (for
example when the flux has dropped below a specified value) or a hard limit on
the length of a ray can be set (for example 2 box lengths).  The impact
parameter for every particle-ray intersection is calculated using a fast AABB
test allowing for the calculation of the photon flux at all the particle-ray
intersections.
 

 

\subsection{Sources}

Any point in the simulation volume can be specified as the beginning of a ray.
Currently, \code is configured to treat point sources whose properties are
specified by an input file, recombination rays from ionized SPH particles, and
background fluxes by specifying points on the simulation volume walls as
sources.

\subsubsection{Diffuse Recombination Radiation}

Here we review the recombination processes that produce ionizing photons in a
H/He gas \citep[e.g.,][]{1989agna.book.....O}.  The following free-bound
transitions produce continuous spectra.
 
\begin{eqnarray}
{\rm HII}  + e \rightarrow {\rm HI(1^2S)}      + \gamma && ({\rm \approx 13.6 \, eV}) \\
{\rm HeII} + e \rightarrow {\rm HeI(1^1S)}     + \gamma && ({\rm \approx 24.6 \, eV}) \\
{\rm HeIII} + e \rightarrow {\rm HeII(2^2S)}   + \gamma && ({\rm \approx 13.6 \, eV}) \\
{\rm HeIII} + e \rightarrow {\rm HeII(2^2P)}   + \gamma && ({\rm \approx 13.6 \, eV}) \\
{\rm HeIII} + e \rightarrow {\rm HeII(1^2S)}   + \gamma && ({\rm \approx 54.4 \, eV}) 
\end{eqnarray}

These spectra can be calculated exactly using the Milne relations.  For Hydrogen, the emission coefficient for the above process is \citep{1989agna.book.....O},  

\begin{equation}
\epsilon_{\rm H}^1 = \frac{2h\nu^3}{c^2} 
\left( \frac{h^2}{2 \pi m_e k_{\rm B} T} \right)^{3/2} \sigma_{\rm H}^1 e^{-h(\nu - \nu_{\rm HI})/k_{\rm B}T} \nHII \nel
\end{equation}
where $\sigma_{\rm H}^1$ is the photo absorption coefficient of Hydrogen in
the ground state.  Similar relations can be derived for the other free-bound
processes, however in practice these highly peaked spectra can be approximated
by delta functions just above the appropriate threshold \footnote{It is a
coincidence that electron captures by HeIII directly to the n=2 level of HeII
have a spectrum peaked at the Hydrogen threshold.} (in parentheses in the
equations above).

Following free-bound captures to excited Helium states, the following
bound-bound transitions can also produce ionizing photons.
\begin{eqnarray}
{\rm HeI(2^3S)} \rightarrow {\rm HeI(1^1S)}    + \gamma && {\rm 19.8 \, eV  } \\
{\rm HeI(2^1P)} \rightarrow {\rm HeI(1^1S)}    + \gamma && {\rm 21.2 \, eV  } \\
{\rm HeI(2^1S)} \rightarrow {\rm HeI(1^1S)}   + 2\gamma && \Sigma {\rm 20.6 \, eV} \\
{\rm HeII(2^2P)} \rightarrow {\rm HeII(1^2S)}  + \gamma && {\rm 40.8 \, eV } \\
{\rm HeII(2^2S)} \rightarrow {\rm HeII(1^2S)} + 2\gamma && \Sigma {\rm 40.8 \, eV } 
\end{eqnarray}
The various bound-bound transitions above have relative probabilities that
depend on the environment (free electron density, temperature, ionization
state) and so the weights to give to these processes should be tailored to
specific applications.

The most straight forward way to deal with this diffuse radiation is to use
the on-the-spot (OTS) approximation.  The OTS approximation makes the
assumption that recombination photons are absorbed in the vicinity (the same
SPH particle) of the point where they are emitted.  Computationally, this
means no ray tracing is necessary for these photons.  For pure Hydrogen
simulations, the OTS approximation amounts to using the reduced recombination
rates in the appendix (case B) \footnote{ Case A rates refer to recombinations
to all atomic levels.  Case B rates refer to recombinations to all but the
first atomic level so that $\alpha_{\rm H}^1 = \alpha_{\rm H}^A - \alpha_{\rm
H}^B$.}.  Here, one is making the assumption that each electron capture
directly to the ground state produces a photon that ionizes a nearby Hydrogen
atom and the two actions effectively cancel one another.

For simulations involving Helium, using the case B rates would be
making the assumption that each HeII recombination to the ground state ionizes
a nearby HeI atom while each HeIII recombination to the ground state ionizes a
nearby HeII atom.  This is a simple first approximation, but some of the
HeII ground state captures will ionize HI, while some of the HeIII ground
state captures will ionize HI and HeI.  To account for this would require a
more detailed adjustment of the recombination and photoionization rates for
the species involved.  The most computationally intensive option is to trace
rays for each of these recombination processes and thereby take account of the
fact that some of the photons will not be absorbed in the SPH particle where
they were created.  Again, the level of detail used should be guided by the
application at hand. With \code it is possible to choose either the 
recombination ray or the OTS approach. 
 We plan to explore the accuracy of the OTS approximation
in various geometries and densities in future work.

\subsection{Optical Depth in SPH}

Ray tracing solutions to the radiative transfer problem solve the equation
along 1D characteristics.  As such, an estimate of the optical depth along
these characteristics is central to the problem.  In the SPH formalism, a
continuous density field is represented by a number of discrete fluid elements
(particles) with smoothing lengths $h_i$.  These smoothing lengths
\footnote{Throughout this work we use the convention that the smoothing kernel
goes to zero at $h$ and not $2h$} are usually defined to keep a constant mass
$M_{\rm sph}$ inside the smoothing volume $ V_i = \frac{4}{3} \pi h_i^3$.  The
properties of the fluid at any point are then estimated by averaging over all
$N$ particles in the simulation weighted by a smoothing kernel.  In practice,
one only averages over nearby particles, but this definition is equally valid
and useful in the derivation to follow.  As an example, the density
$\rho(\mathbf{r}_i)$ at the position $\mathbf{r}_i$ of the $i^{\rm th}$
particle is estimated as,
\begin{equation}
\rho ( {\mathbf r_i} )  \approx \sum_{j=1}^{N} m_j W( | {\mathbf r_i} - {\mathbf r_j} | , h) = 
\sum_{j=1}^{N} m_j  W(r_{ij}  , h)
 \end{equation}
where $m_j$ is the mass of the $j^{\rm th}$ nearest particle and $W(r_{ij},h)$ is a smoothing kernel.  

An estimate of the fluid property need not be made at the position of a
particle.  An averaged value for any fluid property can be defined for an
arbitrary point in space using two techniques.  One is the ``scatter'' method
in which the desired quantity is calculated by averaging over every particle
whose smoothing volume includes the point in question.
\begin{equation}
\rho ( {\mathbf r_i} ) \approx \sum_{j=1}^{N} m_j  W(r_{ij}  , h_j)
 \end{equation}
For this case, a different smoothing length $h_j$ is used in the kernel $W$ for each term.  In the ``gather'' method, a smoothing length is defined for the arbitrary point and the desired quantity is calculated as an average over the particles within this smoothing length.
\begin{equation}
\rho ( {\mathbf r_i} )  \approx \sum_{j=1}^{N} m_j  W(r_{ij}  , h_i)
 \end{equation}
For this case, only one smoothing length $h_i$ is involved.  For definiteness, a popular spline kernel (Monaghan and Lattanzio 1985) is,

\begin{equation} 
W(r,h) = \frac{8}{\pi h^3} \left\lbrace  \begin{array}{ll}
   1 - 6(\frac{r}{h})^2 + 6(\frac{r}{h})^3   &   0 \le r \le \frac{h}{2}  \\ 
   2 ( 1 - \frac{r}{h} )^3           &   \frac{h}{2} < r \le h \\
   0                            &   r > h
 \end{array} \right.
\end{equation}

In our ray tracing scheme, we would like an estimate of the column depth
$N_{\rm cd}$ along a ray that has intersected a number of SPH particles.
Formally this is,
\begin{equation}
N_{\rm cd} = \int_0^L \rho ({\mathbf r}) dl \approx \int_0^L \sum_{j=1}^{N} m_j  W(r_{lj}  , h_j) dl
\end{equation}
where the scatter interpretation has been used.  The summation extends over
all particles and $l$ parameterizes the distance along the ray.  Interchanging
the order of integration and summation we have,
\begin{equation}
N_{\rm cd} \approx \sum_{j=1}^{N} \int_0^L m_j  W(r_{lj}  , h_j) dl
\end{equation}

This amounts to a line integral through the smoothing kernel of each particle
whose smoothing volume is pierced by the ray.  This integral is calculated by
tabulating it as a function of impact parameter $b$ for one value of $h$,
namely $h=1$.  We can recover the line integral for any $b$ and $h$ through a
rescaling of the tabulated value.  This technique delivers the optical depth
to a point along the ray, as well as the contribution from each particle in
the raylist to that optical depth.

There are two related approximations that come into play here.  The first
involves the reordering of the terms so that those involving the same particle
are adjacent.  This is valid as long as the density doesn't vary much within a
smoothing length which is true by construction.  The next involves the
endpoint of the ray.  Given this reordering, the contribution to the density 
(at the terminus of the ray) from particles which have centers further down the ray but smoothing 
volumes' that contain the end point, will be unaccounted for.
This is a small correction for the reason given above, and the fact that the
photons are usually nearly all absorbed when the ray is terminated.

\subsection{Photon Packets}

In our Monte Carlo method, the radiation field is discretized into photon
packets and transported along rays.  Each packet contains a large number of
monochromatic photons sampled from an arbitrary spectral energy distribution
(SED).  The direction along which a photon packet is transported is also
sampled from an emission profile distribution.  This is true whether the
packet represents emission from a point source, diffuse recombination
emission, or background radiation originating outside the computational
volume.  The starting locations for rays can be any point within the
computational volume including points on the faces of the simulation volume.

For each ray that is cast, a source is selected at random weighted by its
luminosity.  This ensures a population of photon packets with roughly the same
energy as opposed to the same number of photon packets being traced from each
source regardless of their luminosities.

The base resolution of a simulation is determined by how many SPH particles 
$N_{\rm p}$ are used to sample the continuous density field.  The degree to which 
the sampling of the radiation field approaches the base resolution is determined 
by the number of rays traced $N_{\rm r}$.  For isotropic sources and
homogeneous density fields, the average number of particle intersections per
ray is $\approx$ $N_{\rm p}^{1/3}$.  It follows that the total number of
intersections is on the order of $N_{\rm r} N_{\rm p}^{1/3}$ and
that the average number of intersections per particle is $N_{\rm r} N_{\rm p}^{-2/3}$.
This gives a rough estimate of how many times the radiation field is sampled
at each particle.  In practice, particles closer to sources will be sampled more often and 
it is better to conduct a convergence study than to rely on pre-calculated estimates of resolution.

\subsection{Particle-Ray Intersections}

Once a photon packet has been constructed, it is propagated along a ray.
\code uses a data object called a raylist to store the intersections of a ray
drawn from the source, and all the SPH particles whose smoothing volumes are
pierced by the ray.  This can be done using vacuum, or periodic
boundary conditions (in the latter case a maximum length must be
specified).  The search for these intersections needs to be as efficient as possible.

\begin{figure}
\centerline{
\psfig{file=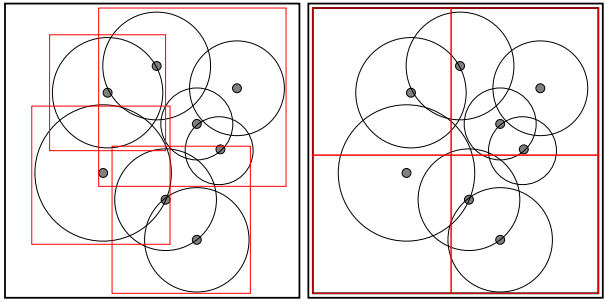,angle=-90.,width=8.0truecm}
}
\caption{Cartoon 2D representation of several SPH particles.  The left panel 
shows the positions of the axis aligned bounding boxes and the right panel 
indicates the quad-tree cell boundaries.
\label{aabbfig}
}
\end{figure}

Because the smoothing lengths within a cosmological box can vary by more then three
orders of magnitude, we organize the particles into an oct-tree.  This manner of
storing particles is common and many SPH codes produce an oct-tree during the
course of hydrodynamic calculations.  Our code could be trivially modified to
use a pre-constructed tree although currently \code constructs its own for each
density field it ray traces.  We augment the standard oct-tree by associating
with every cell an axis-aligned bounding box (AABB) that is just large enough
to encompass the smoothing lengths of all the particles in that cell (Figure
\ref{aabbfig}).

Each cell of our tree contains either particles or daughter cells.  The
maximum number of particles in a leaf is specified in a configuration file.
We have found 12 to be a reasonable choice.  The search for particle-ray
intersections proceeds exactly as in the case of a simple SPH neighbor
search.  Starting with the root cell, the AABB of the cells are tested for
intersection with the ray.  In case of intersection, the cell is opened and
the search proceeds on the daughter cells.  This process continues until a
leaf with no further refinements is encountered, in which case the particles
in the leaf are tested to see which of them are intersected by the ray.  This
also produces the impact parameters of each particle in the raylist.
 
The intersection test of the ray with the AABB is done using pl\"{u}cker
coordinates \citep{mahovsky:plu04}, a very fast method used in computer
graphics. This method tests the ray against the edges comprising the
silhouette of the AABB instead of testing against the individual faces, it is
division free and consists of a number of simple dot-product operations.

\subsection{Solvers}

Once we have a photon packet and a list of the ray-particle intersections
stored in a raylist, we can proceed to update all the particles in the
raylist.  \code offers two choices for this task.  The first is an adaptive
Runge-Kutta (RK) method.  A set of formulas due to \cite{FehlbergRK} provide
solutions that are accurate to fifth order in the time step. Step size
control is provided 
using the truncation error as estimated by the embedded fourth order
solution.  The Cash-Karp coefficients \cite{CashKarpRK} are used to take the
variable length time steps.  This option is included because it is simple and
could be easily modified if a user needs to add extra physics into the
solution routine.  It is also guaranteed to conserve photons to an arbitrary
accuracy by forcing the number of ionizations in a particle to equal the
number of photons removed from a packet.

The second solution method is an iterative solver based on time averaged
photoionization rates and optical depths.  The time averaging removes the need
for very short time steps, but requires approximate analytic solutions and in
the case of extremely long time steps will not conserve photons exactly.  The
main advantage is that the number of iterations necessary to obtain a
converged solution can be much smaller than the number of RK time steps
necessary to obtain the same solution.  This method was introduced in
\cite{2006NewA...11..374M} and a more detailed description of it can be found
there.

\subsubsection{Runge-Kutta}

For each intersection we determine the time $t_{\rm li}$ since the particle
has last been intersected by a ray.  The photon flux $\dot{N}_{\gamma}$ at the
particle is estimated as the photons left in the ray $N_{l}$ divided by
$t_{\rm li}$.  This is taken to be the first guess for a time step in the
solution of the system of coupled differential equations (Eqs. 15,17,18, and
52).  The optical depth through the particle is estimated using the technique
described in \S 3.2.  This allows the calculation of the total photoionization
rate $\Gamma$ which along with the values of $G_{A}$ and $R_{I}$ 
are all that is
necessary to calculate the right hand sides of the equations mentioned above.
The photoionization rate is 

\begin{equation}
\Gamma = \dot{N}_{\gamma} \left( 1 - e^{- \Delta \tau} \right)  \times 
\frac{m_{\rm H}}{M_p} \left[ X \xHI + \frac{Y}{4} (\xHeI + \xHeII)  \right]^{-1} 
\end{equation}

where $\dot{N}_{\gamma}$ is the photon flux at the particle, $\Delta \tau$ is
the optical depth through the particle, $m_{\rm H}$ is the mass of a Hydrogen
atom, $M_p$ is the mass of the particle, $X$ is the Hydrogen mass fraction and
$Y$ is the Helium mass fraction.  Here, $\xHeI$ and $\xHeII$
should be set to zero for frequencies less than their respective thresholds.
$\Gamma$ for the individual species is calculated from the ratio of their
optical depths to the total optical depth.

\begin{equation} 
\Gamma_A = \frac{\Delta \tau_A}{\Delta \tau}
\end{equation}

The number of photons absorbed by a particle $N_{\rm a}$ is obtained
by solving an extra
differential equation, as
the photoionization rate is allowed to vary for each
substep that the RK routine takes.

The photon packet is followed along a ray until the fraction of photons left
is below a threshold or until the photon packet has reached a pre-determined
distance along the ray.  A typical value for the photon tolerance is $1.0
\times 10^{-10}$ of the initial photons in the packet.  This determines the
level of photon conservation and can be set arbitrarily low.

\subsubsection{Iterative}

If the iterative solution method is chosen, the default time step $t_{\rm li}$
can be used for most updates.  The first step in this solution method is to
initialize the time averaged ionization fractions to the current values in a
particle.  These are used to make a first guess at the time averaged optical
depth and photoionization rate.

\begin{eqnarray} 
\left\langle \Gamma \right\rangle &=& \dot{N}_{\gamma} \left( 1 - e^{- \left\langle \Delta \tau \right\rangle} \right)  \times  \nonumber \\
&& \frac{m_{\rm H}}{M_p} \left[ X \left\langle \xHI \right\rangle + 
\frac{Y}{4} (\left\langle \xHeI \right\rangle + \left\langle \xHeII \right\rangle)  \right]^{-1} 
\end{eqnarray} 

This time averaged photoionization rate is used to calculate the time averaged
ionization fractions (Eqs. 52-56) which are in turn used to update the time
averaged optical depths.  The optical depths can then be used to find a new
photoionization rate and the iteration proceeds until we have reached
convergence in the electron number density and temperature.  Because the
heating and cooling rates are themselves functions of temperature (as opposed
to the recombination and collisional ionization rates which are not functions
of the ionization fraction), the temperature is always updated using the RK
routine and the ionization state at each iteration.  Once the iterations have
converged, the final state of the particle is calculated using eqs. 28 - 32.

\section{Code Verification}

Here we present the results of \code on the tests outlined in the radiative
transfer comparison project paper by \cite{2006MNRAS.371.1057I}.  We will
describe the tests briefly, but refer the reader to the reference for details.
For tests 1 and 2, the initial conditions were set up using SPH particles that
were evolved into a glass state using the code {\small GADGET-2}
\citep{2005MNRAS.364.1105S}.  The distribution of particles extended two
smoothing lengths further then the required box size in order to avoid edge
effects in the density field and vacuum boundary conditions were used.

Tests 1 and 2 contained exactly $128^3$ SPH particles within the $6.6^3$
kpc$^3$ volume while test 3 contained 2,135,842 particles in this same volume.
For the third test, 853,442 particles were first evolved into a glass and then
SPH particles were randomly placed inside the clump until the density there had
reached 200 times the density outside the clump.  The initial conditions for
the cosmological test are discussed in \S 4.1.4. The SPH density field as well
as the temperature and ionization fraction variables have been interpolated
onto a $128^3$ grid to make the surface plots and for submission to the
Comparison Project website and so for those figures we will refer to grid
cells.

\subsection{Test 1. Pure hydrogen isothermal H II region expansion}

The first test considers the growth of a Str{\"o}mgren sphere in a uniform density
field consisting of pure hydrogen.  The source, placed in the corner of a
$6.6$ kpc box, emits $ \dot{N}_\gamma = 5.0 \times 10^{48}$ Rydberg photons
per second.  The density of hydrogen is $n_{\rm H} = 1.0 \times 10^{-3} {\rm
cm}^{-3}$, and the temperature is fixed at $T=10,000$ K.  The system is
evolved for $500$ Myr or approximately four recombination times.  The state of
the system is examined at 10 and 100 Myr when the I-front is growing quickly and at
500 Myr when the photoionizations and recombinations have balanced and the HII
region has reached its final Str\"omgren radius.

\begin{figure*}
\begin{center}
\psfig{file=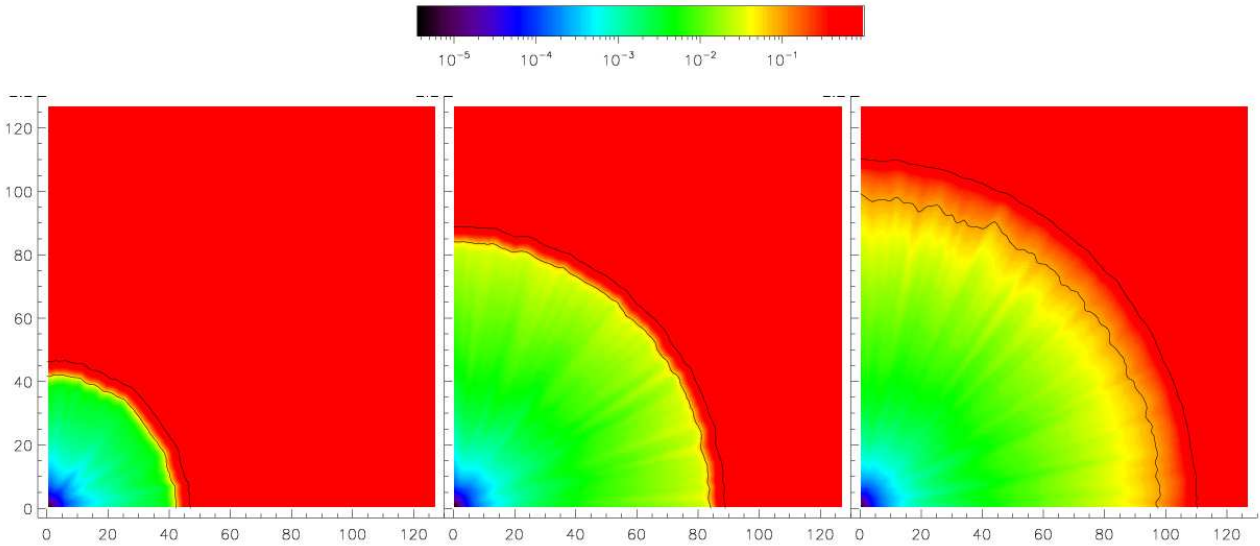,angle=0.,width=18.0truecm}
\end{center}
\caption{ \testone  Surface plot of $\xHI$ cut through the simulation volume at coordinate z = 0 at t = 10 (left), 30 (middle), and 500 (right) Myr.  The contours are at $ \xHI = 0.1$ and $ \xHI = 0.9 $.  The number of rays traced  $N_{\rm t}$ by \code is $q \times 10^8$ where $q = t/500$ is the fraction of elapsed time.  The axes are measured in grid cells. }
\label{T1xsurf135}
\end{figure*}

\begin{figure*}
\begin{center}
\psfig{file=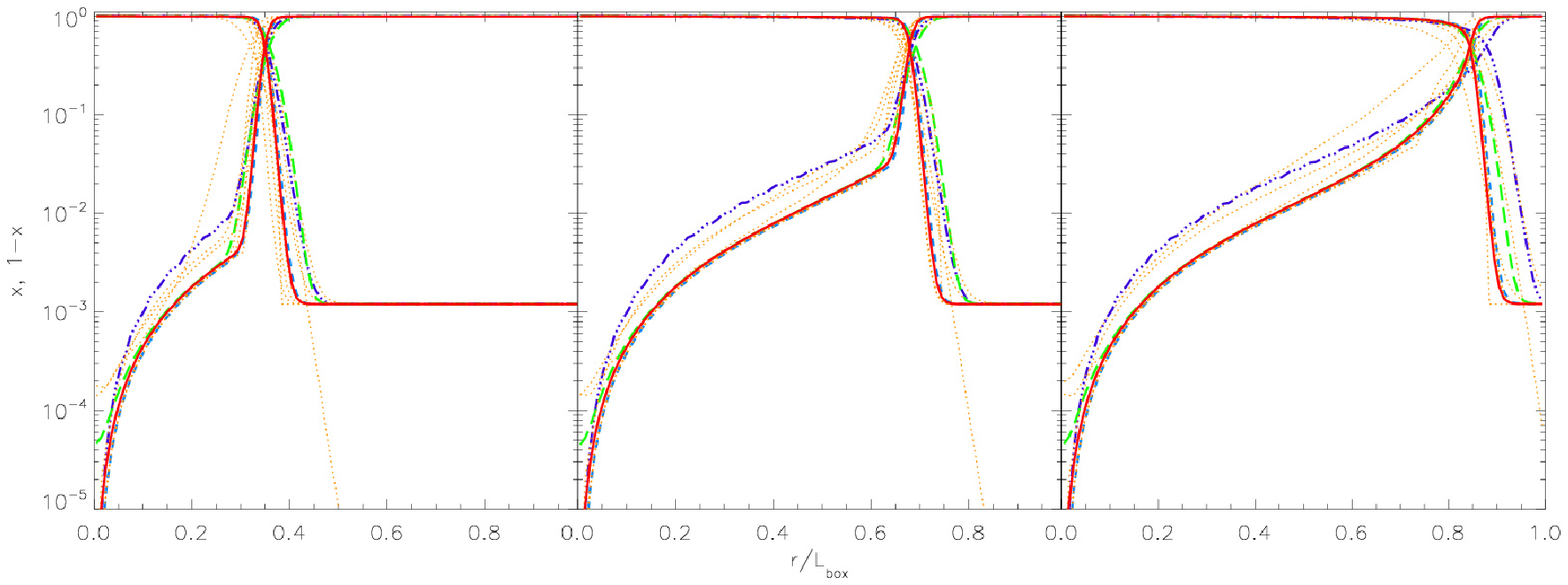,angle=0.,width=18.0truecm}
\end{center}
\caption{\testone  $\xHI$ and $\xHII$ radial profiles at t = 10 (left), 100 (middle), and 500 (right) Myr.  The number of rays traced  $N_{\rm t}$ by \code is $q \times 10^7$ where $q = t/500$ is the fraction of elapsed time.  \figcapt }
\label{T1xvr135}
\end{figure*}

This simple test has the advantage that numerical results can be compared
directly with an analytic solution for the position and velocity of the
I-front versus time.  \code finds agreement with these solutions at the
several percent level (Figure \ref{T1xvr135}).
The analytic front width, defined as the
distance over which the neutral fraction goes from $\xHI = 0.1$ to $\xHI = 0.9
$ is $ r_{\rm if} \approx 18 \lambda_{\rm mfp} \approx $ 14 grid cells
\citep{2006MNRAS.371.1057I} and is also reproduced by \code.  This can be seen
in figure \ref{T1xsurf135}, by noting that the contours are at $\xHI = 0.1$
and $\xHI = 0.9$.

\begin{figure*}
\begin{center}
\psfig{file=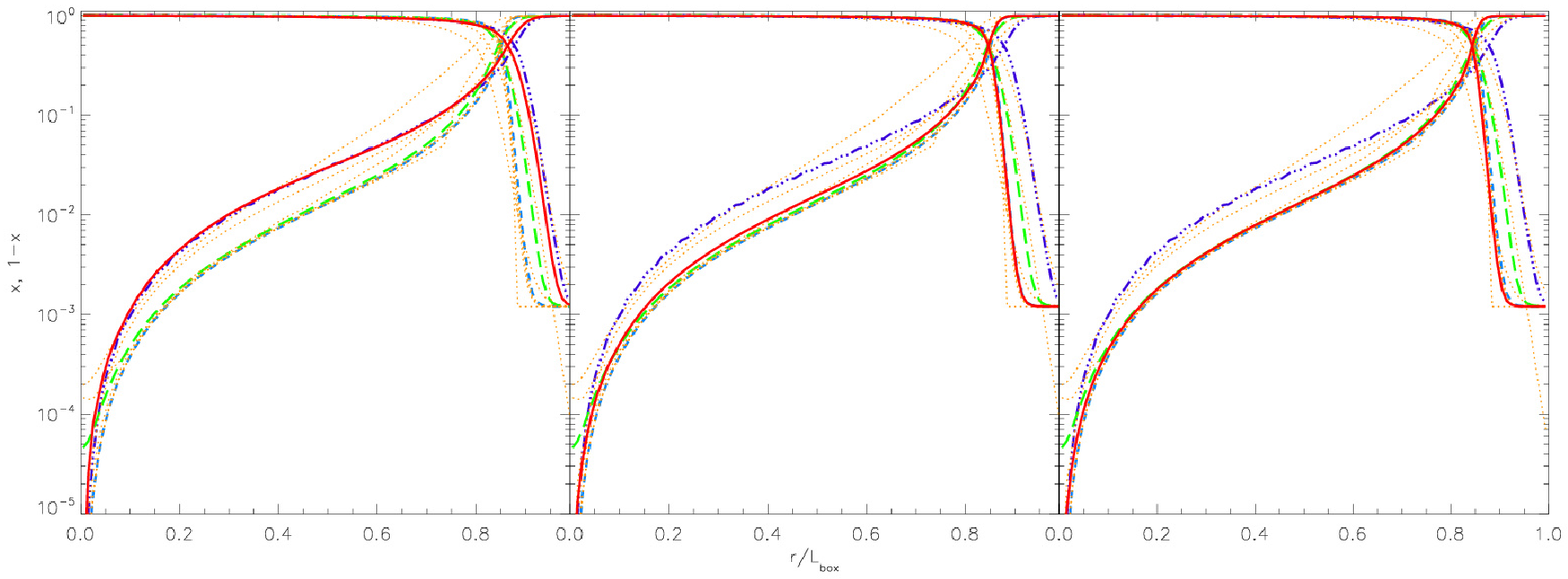,angle=0.,width=18.0truecm}
\end{center}
\caption{\testone  $\xHI$ and $\xHII$ radial profiles at t = 10 Myr with $N_{\rm t}= q \times 10^6$ (left), $N_{\rm t}= q \times 10^7$ (middle), and $N_{\rm t}= q \times 10^8$ (right) where $N_{\rm t}$ is the number of rays traced and $q = 10/500$ is the fraction of elapsed time. \figcapt }
\label{T1xvr1_r678}
\end{figure*}

The size of the highly ionized proximity region produced by all the codes in
the comparison project can be seen by examining the neutral fraction lines in
figure \ref{T1xvr135}.  Our solution follows closely that produced by {\small
RSPH}.  There are several codes that produce slightly smaller proximity
regions then the rest.  These codes may not have converged or be unable to
reach the native resolution of the density field as \code produced similar results before converging as can be seen in figure \ref{T1xvr1_r678}.

The anisotropies seen in the surface plots of \code's results have three major
sources: the Monte Carlo method used, Poisson fluctuations in the sampling of
solid angle by the rays, and fluctuations in the density field due to the SPH
glass.  In future work, it may be useful to
use a low discrepancy sequence instead of
uniformly distributed pseudo random numbers to generate ray directions,
however given the fact that these anisotropies can be reduced by tracing more
rays and the agreement of the radially averaged profiles we consider this a
minor problem.

\subsection{Test 2. HII region expansion: the temperature state}

Test 2 is identical to Test 1 except the gas temperature is now allowed to
 vary, starting from
 an initial value of $T=100$K and the spectrum of the ionizing source is
taken to be that of a $T = 10^5$ K blackbody.  This is a more demanding test
as the heating times are very short in comparison with the other
characteristic times and now the rays must sample frequency space as well.
Multi-frequency photon packets would alleviate the need for this extra
sampling and are planned as an additional
option in future versions of \code.

\begin{figure*}
\begin{center}
\psfig{file=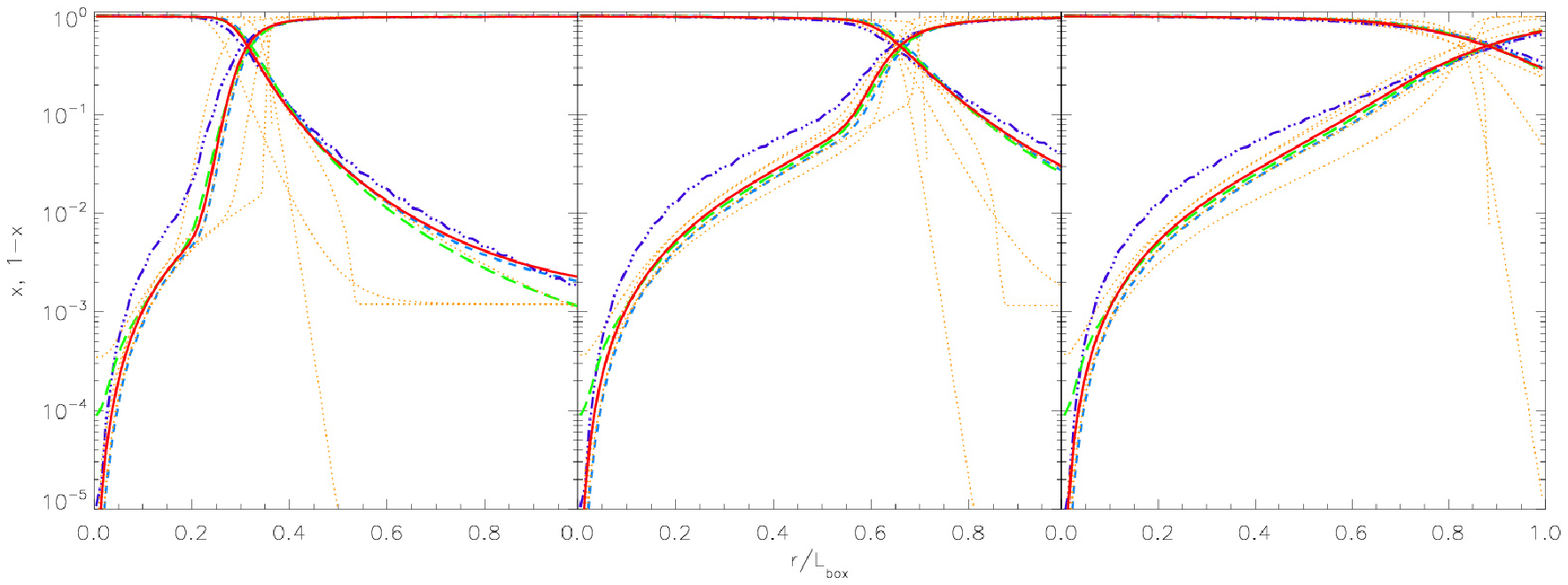,angle=0.,width=18.0truecm}
\end{center}
\caption{\testtwo  Spherically averaged $\xHI$ and $\xHII$ profiles at t = 10 (left), 100 (middle), and 500 (right) Myr.  The number of rays traced  $N_{\rm t}$ by \code is $q \times 10^7$ where $q = t/500$ is the fraction of elapsed time.  \figcapt }
\label{T2xvr135}
\end{figure*}

\begin{figure*}
\begin{center}
\psfig{file=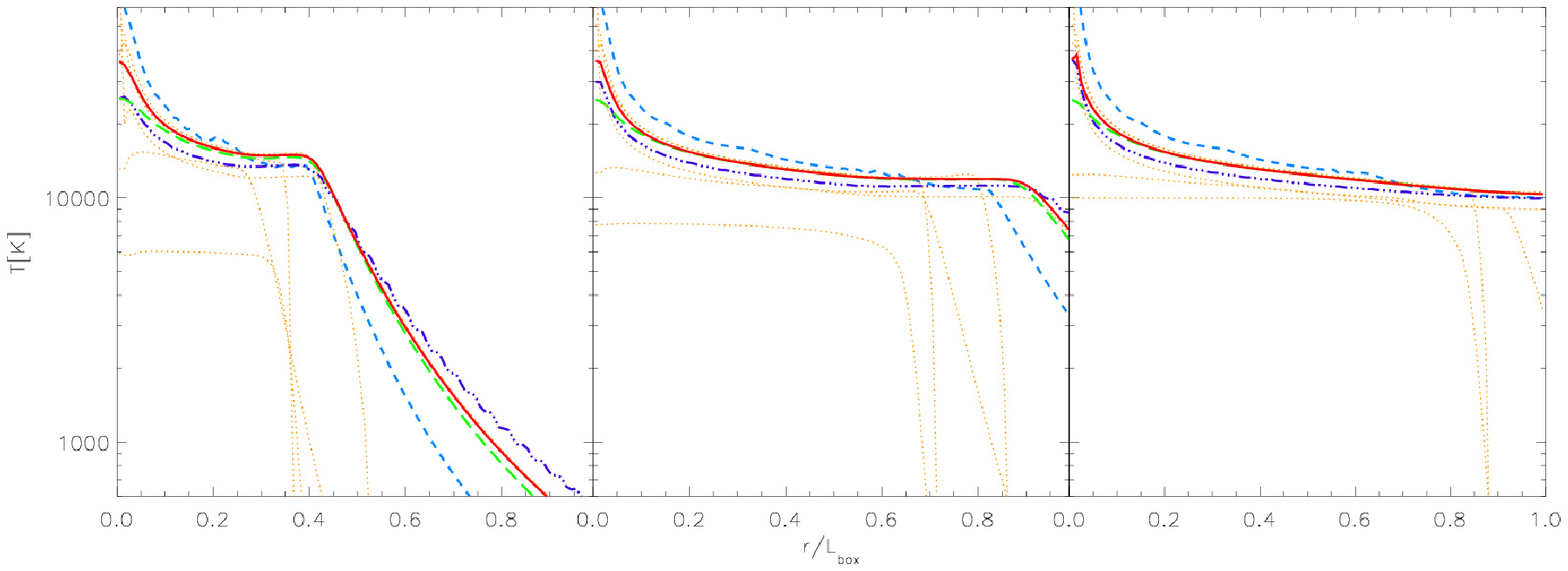,angle=0.,width=18.0truecm}
\end{center}
\caption{\testtwo.  Spherically averaged temperature profiles at t = 10 (left), 100 (middle), and 500 (right) Myr.  The number of rays traced  $N_{\rm t}$ by \code is $q \times 10^7$ where $q = t/500$ is the fraction of elapsed time.  \figcapt }
\label{T2tvr135}
\end{figure*}

The radially averaged ionization ( figure \ref{T2xvr135} ) and temperature (
figure \ref{T2tvr135} ) profiles that \code produced for this test again
follow very closely those produced by {\small RSPH}.  \code correctly produces
a preheated region ahead of the ionization front due to the accurate treatment
of high energy photons with long mean free paths and a large amount of energy
to deposit as heat.


\begin{figure*}
\begin{center}
\psfig{file=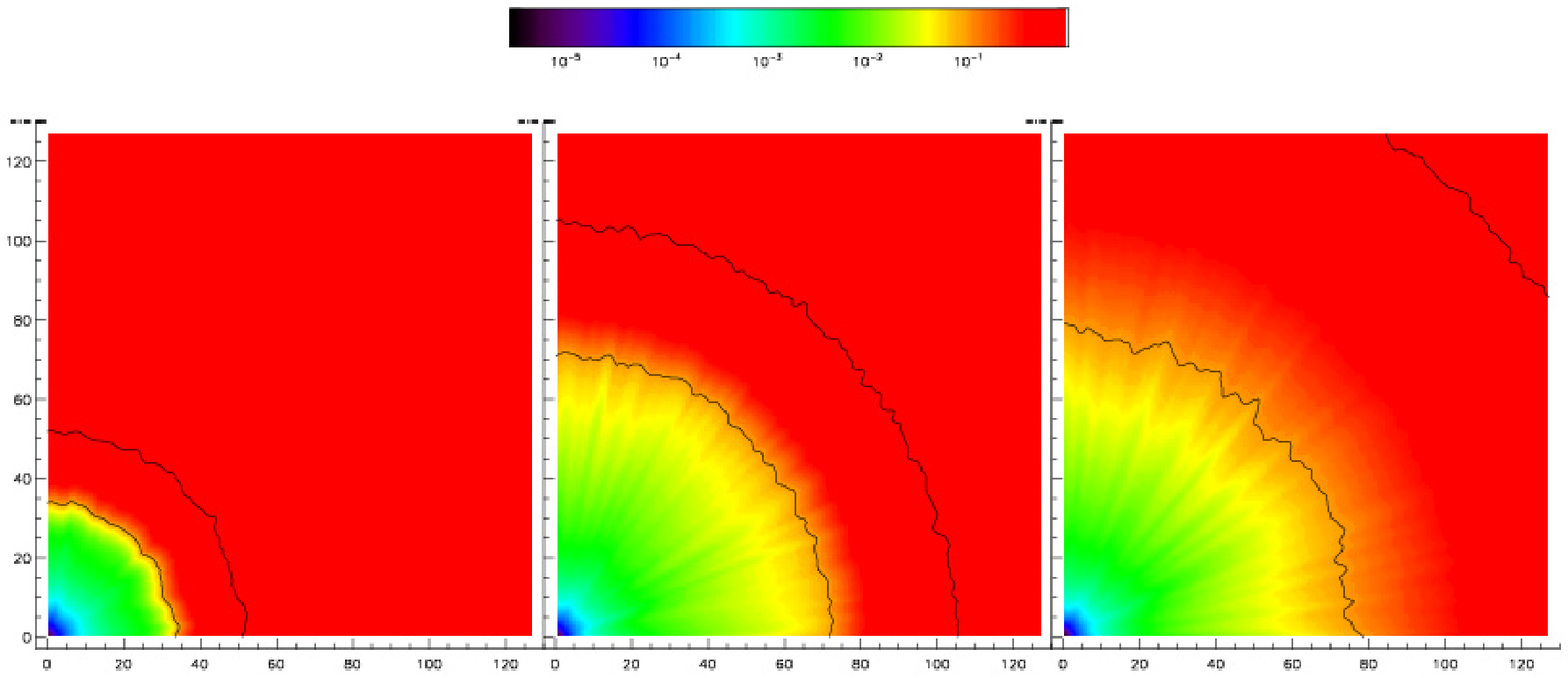,angle=0.,width=18.0truecm}
\end{center}
\caption{\testtwo Surface plot of $\xHI$ cut through the simulation volume at
coordinate z = 0 and t = 10 (left), 100 (middle), and 500 (right) Myr.  The
contours are at $ \xHI = 0.1$ and $ \xHI = 0.9 $.  The number of rays traced  $N_{\rm t}$ by \code is $q \times 10^8$ where $q = t/500$ is the fraction of elapsed time.  The axes are measured in
grid cells.}
\label{T2xsurf135}
\end{figure*}

\begin{figure*}
\begin{center}
\psfig{file=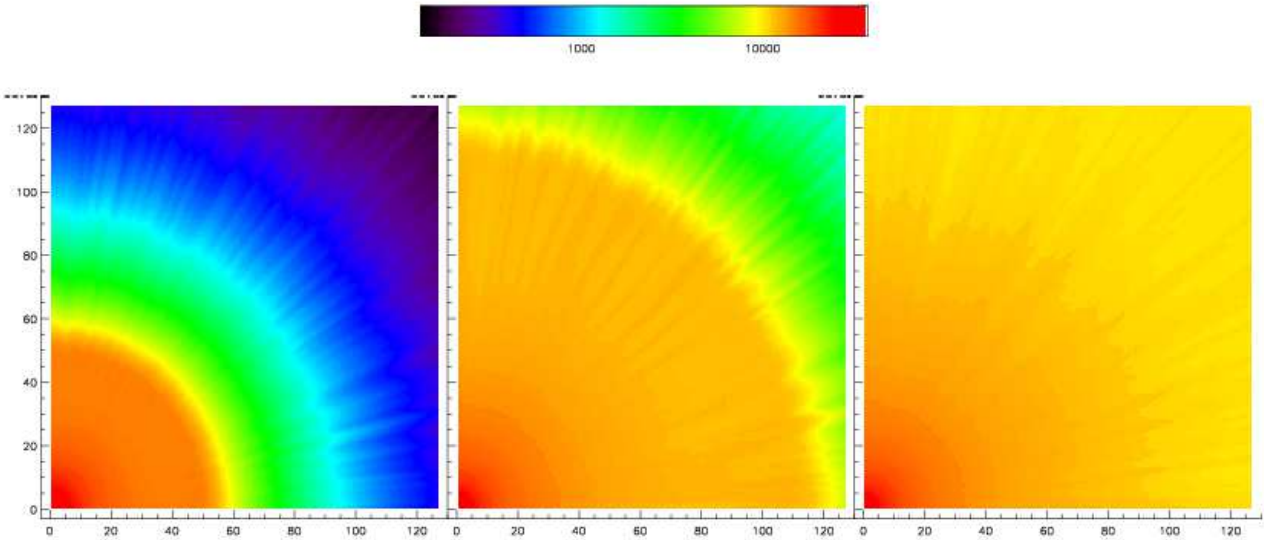,angle=0.,width=18.0truecm}
\end{center}
\caption{\testtwo  Surface plot of the temperature cut through the simulation volume at coordinate z = 0 and t = 10 (left), 100 (middle), and 500 (right) Myr.  The number of rays traced  $N_{\rm t}$ by \code is $q \times 10^8$ where $q = t/500$ is the fraction of elapsed time.  The axes are measured in grid cells.}
\label{T2tsurf135}
\end{figure*}

\subsection{Test 3. I-front trapping in a dense clump and the formation of a shadow}

In this test, a cold dense spherical clump of hydrogen
gas is embedded in a hot diffuse
background.  The dimensions of the simulation box are the same as above
(6.6$^3$ kpc$^3$).  The clump has a radius $r_c = 0.8$ kpc and its center is
located at, $x_c = (5.0,3.3,3.3)$ kpc. The density contrast is $n_{\rm
in}/n_{\rm out} = 200$ with $n_{\rm out} = 2 \times 10^{-4} $cm$^{-3}$.  The
gas in the clump is set to a temperature of $T_{\rm in} = 40$ K while the gas
outside the clump is initialized to $T_{\rm out} = 8000$ K.  The entire
$x=0$ side of
the simulation box is taken to be
a $T=10,000$K blackbody source with a constant
photon flux, $F = 10^6$ s$^{-1}$ cm$^{-2}$ into the box.  The test is designed
so that the initially fast moving ionization front will be trapped in the
clump due to the higher recombination rate there.

\begin{figure*}
\begin{center}
\psfig{file=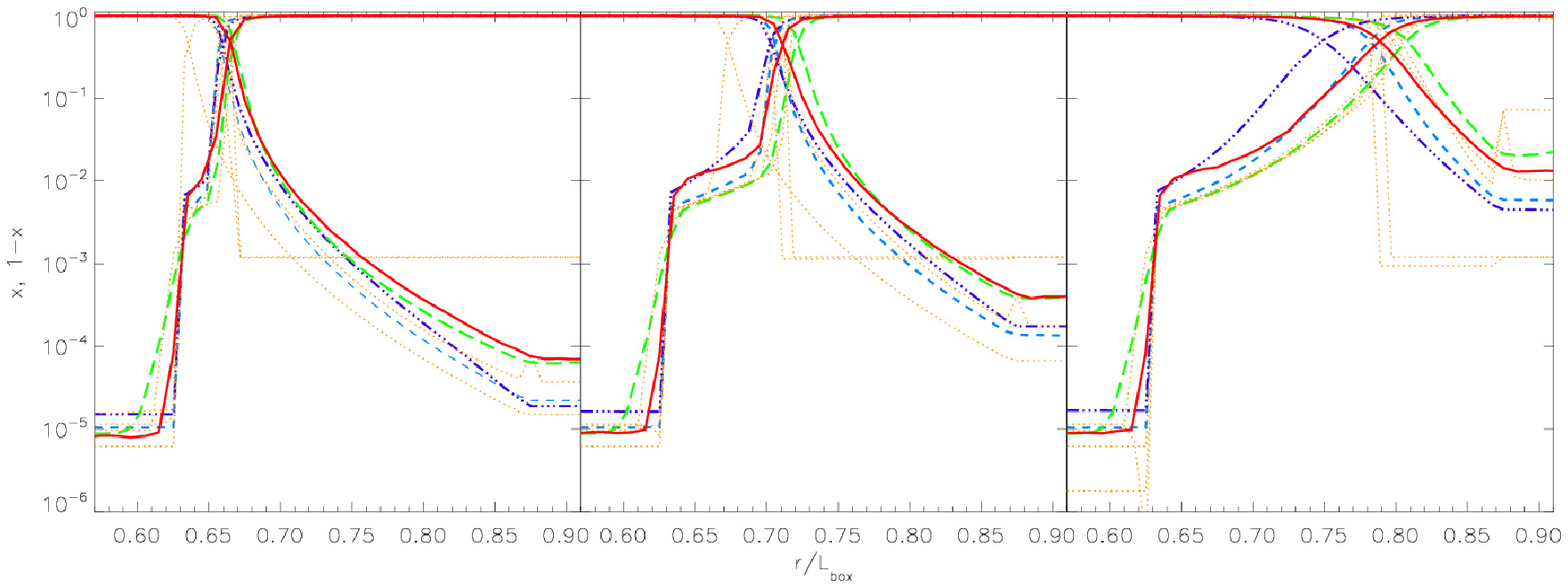,angle=0.,width=18.0truecm}
\end{center}
\caption{\testthree  $\xHI$ and $\xHII$ profiles along the axis of symmetry at t = 1 (left), 3 (middle) , and 15 (right) Myr.  The number of rays traced  $N_{\rm t}$ by \code is $q \times 10^8$ where $q = t/15$ is the fraction of elapsed time.  \figcapt }
\label{T3xvr135}
\end{figure*}

\begin{figure*}
\begin{center}
\psfig{file=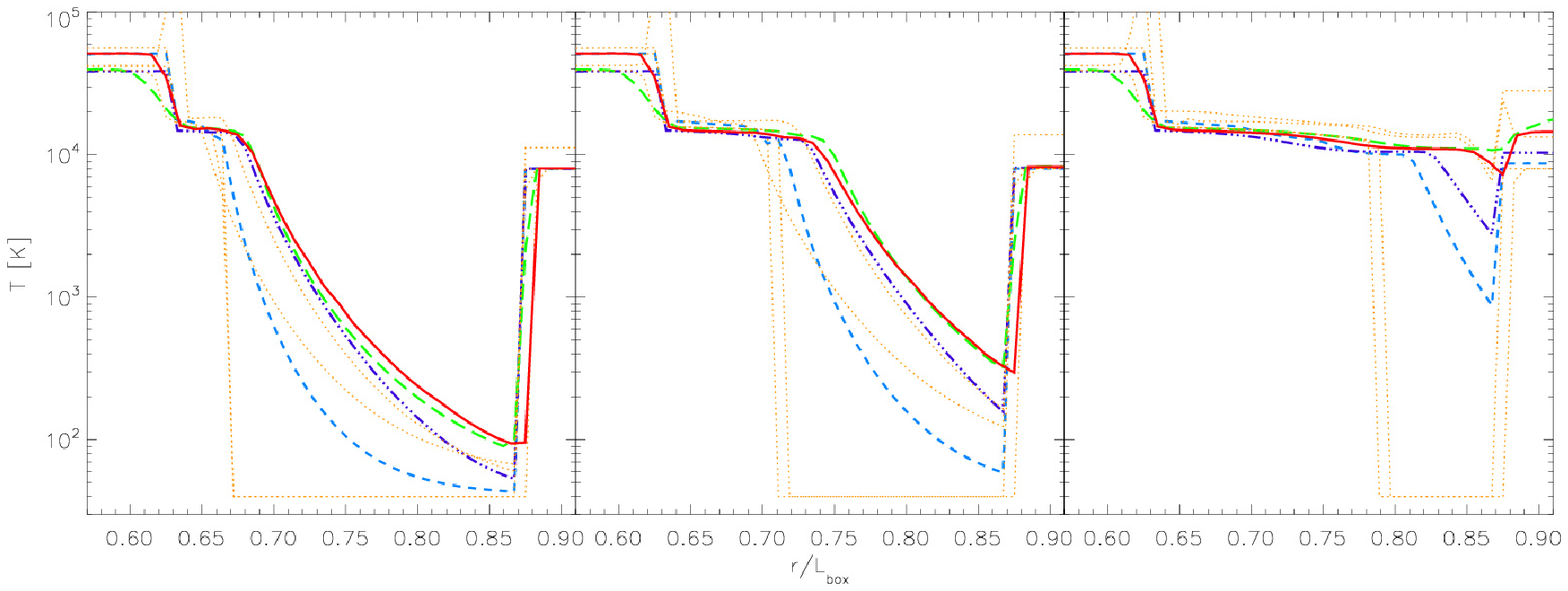,angle=0.,width=18.0truecm}
\end{center}
\caption{\testthree  Temperature profiles along the axis of symmetry at t = 1 (left), 3 (middle), and 15 (right) Myr.  The number of rays traced  $N_{\rm t}$ by \code is $q \times 10^8$ where $q = t/15$ is the fraction of elapsed time.  \figcapt }
\label{T3tvr135}
\end{figure*}

In figures \ref{T3xvr135} and \ref{T3tvr135} we show the ionization and
temperature profiles along a small cylinder through the axis of symmetry.  For
the Comparison Project grid data, we used the four central columns of grid
cells (where each grid cell is $\approx$ 0.05 kpc in length) and for our SPH
data we used all the particles whose centers lie within 0.05 kpc of the axis
of symmetry.  The variation between all the codes is larger for this test than the first two.
\code produces results that lie between those of {\small RSPH} and {\small CRASH} 
for the ionization profiles and results that follow {\small RSPH} for the temperature profiles.

\begin{figure*}
\begin{center}
\psfig{file=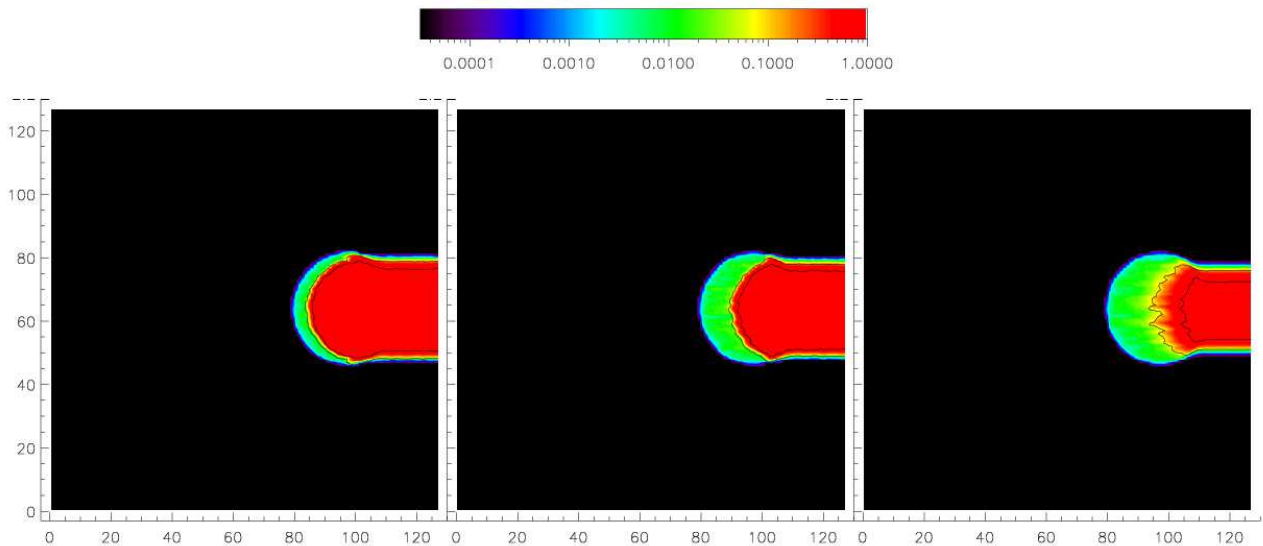,angle=0.,width=18.0truecm}
\end{center}
\caption{\testthree  Surface cut of the neutral fraction at t = 1 (left), 3 (middle), and 15 (right) Myr.  The number of rays traced  $N_{\rm t}$ by \code is $q \times 10^8$ where $q = t/15$ is the fraction of elapsed time.  Shown are the results from \code }
\label{T3xsurf135}
\end{figure*}

\begin{figure*}
\begin{center}
\psfig{file=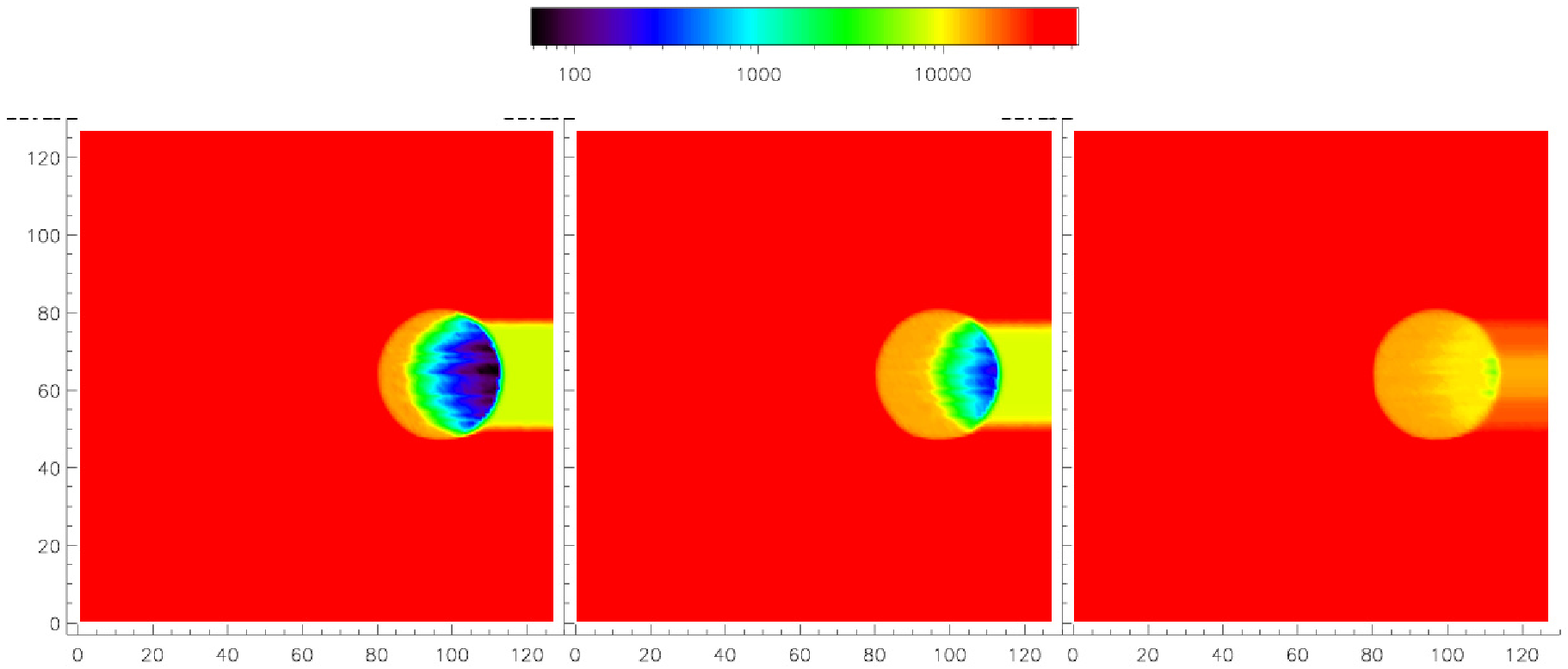,angle=0.,width=18.0truecm}
\end{center}
\caption{\testthree  Surface cut of the temperature at t = 1 (left), 3 (middle), and 15 (right) Myr.  The number of rays traced  $N_{\rm t}$ by \code is $q \times 10^8$ where $q = t/15$ is the fraction of elapsed time.  Shown are the results from \code }
\label{T3tsurf135}
\end{figure*}

This test features many of the effects that contribute to a reprocessing of
the intergalactic background radiation field.  Specifically, self shielding,
shadowing, and spectral hardening.  \code produces an ionization front that
moves quickly to the clump and is trapped near the center as it should be.
The differences with the other codes mostly have to do with the amount of
shadowing and self shielding (ionization and temperature).  The high energy
photons that are sampled by \code produce slightly weaker shadows directly
behind the clump than most of the other codes.  The results are much more
pronounced for the temperature than they are for ionization.  In the last
panel of figure \ref{T3xsurf135} the self shielded section of the clump and
the area in the shadow of the clump are only slightly ionized, however in the
last panel of figure \ref{T3tsurf135} the high energy photons have raised the
temperature of the whole initially cool clump to $\approx 10,000$ K and begun
to heat the shadowed region behind it above its initial temperature of $8,000$
K,

\subsection{Test 4. Multiple sources in a cosmological density field}

The most realistic test preformed in the RT comparison project involved
sources placed in the 16 most massive halos of a cosmological
simulation snapshot at z=9.  The box size is 0.5 $h^{-1}$ comoving Mpc and the gas was initially neutral with a temperature of 100 K.  The sources were assumed to emit $f_{\gamma} = 250$ photons per atom over $t_s$ = 3 Myr leading to a photon flux $\dot{N}_{\gamma}$ of,
\begin{equation}
\dot{N}_{\gamma} = f_{\gamma} \frac{M \Omega_b}{\Omega_0 m_{\rm H} t_s}. 
\end{equation}
where $M$ is the halo mass.  The system was then evolved for 0.4 Myr.

In order for \code to complete this test it was necessary to convert the
density field data  from a grid based
representation  into an SPH density field.  This was accomplished
using the following procedure: we start from a smooth glass distribution
representing a constant density field equal to the peak density $\rho_{\rm
max}$ of the grid data. Particles are then selected for removal according to a comparison
of the density $\rho_p$ at the particle position with a random density
$\rho_X$ where $0<\rho_X<\rho_{\rm max}$. This gives a density field which is
smoother in high density regions than it would be if a normal
acceptance-rejection test with random particle locations was used.  In low
density regions the noise is higher because the distribution becomes less
glass like as particles are removed.  The result was a close approximation to
the gridded density field using 1,957,344 SPH particles.  The test does show
some extraneous noise from the conversion to particles, but this noise is
smaller than the difference between different methods so for our comparison,
the simple procedure outlined above was good enough.  \code is intended 
to be run on density fields produced by SPH hydrodynamic simulations in which this sort of noise would not occur.

\begin{figure}
\begin{center}
\psfig{file=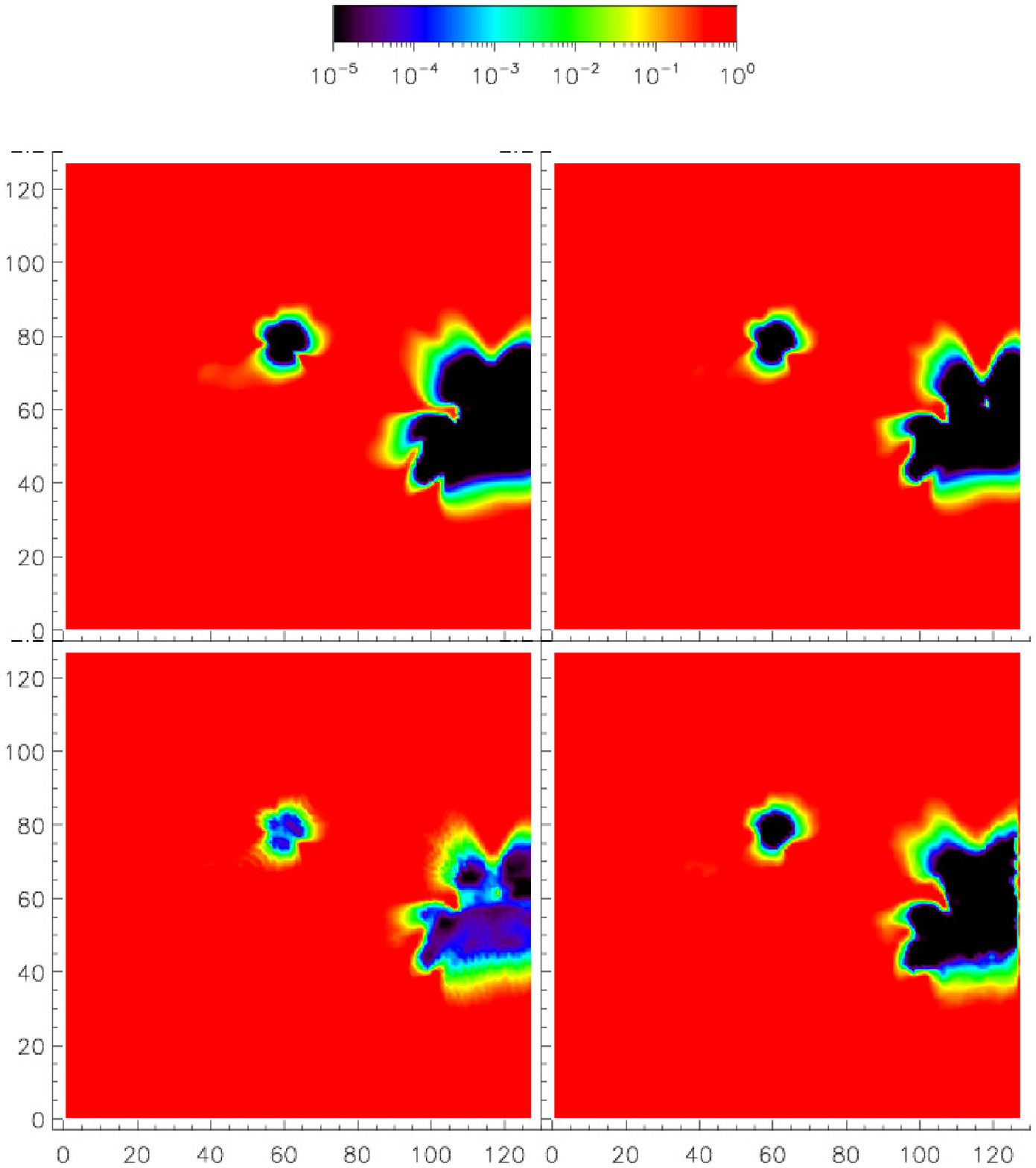,angle=0.,width=9.0truecm}
\end{center}
\caption{\testfour  Surface cut of the neutral fraction through the middle of the simulation volume at t = 0.05 Myr.  The number of rays traced  $N_{\rm t}$ by \code is $q \times 10^8$ where $q = t/0.4$ is the fraction of elapsed time.  Beginning in the top left corner and proceeding clockwise are the results from {\small C$^2$-Ray}, {\small FTTE}, \code, and {\small CRASH}.}
\label{T4x1surf}
\end{figure}
\begin{figure}
\begin{center}
\psfig{file=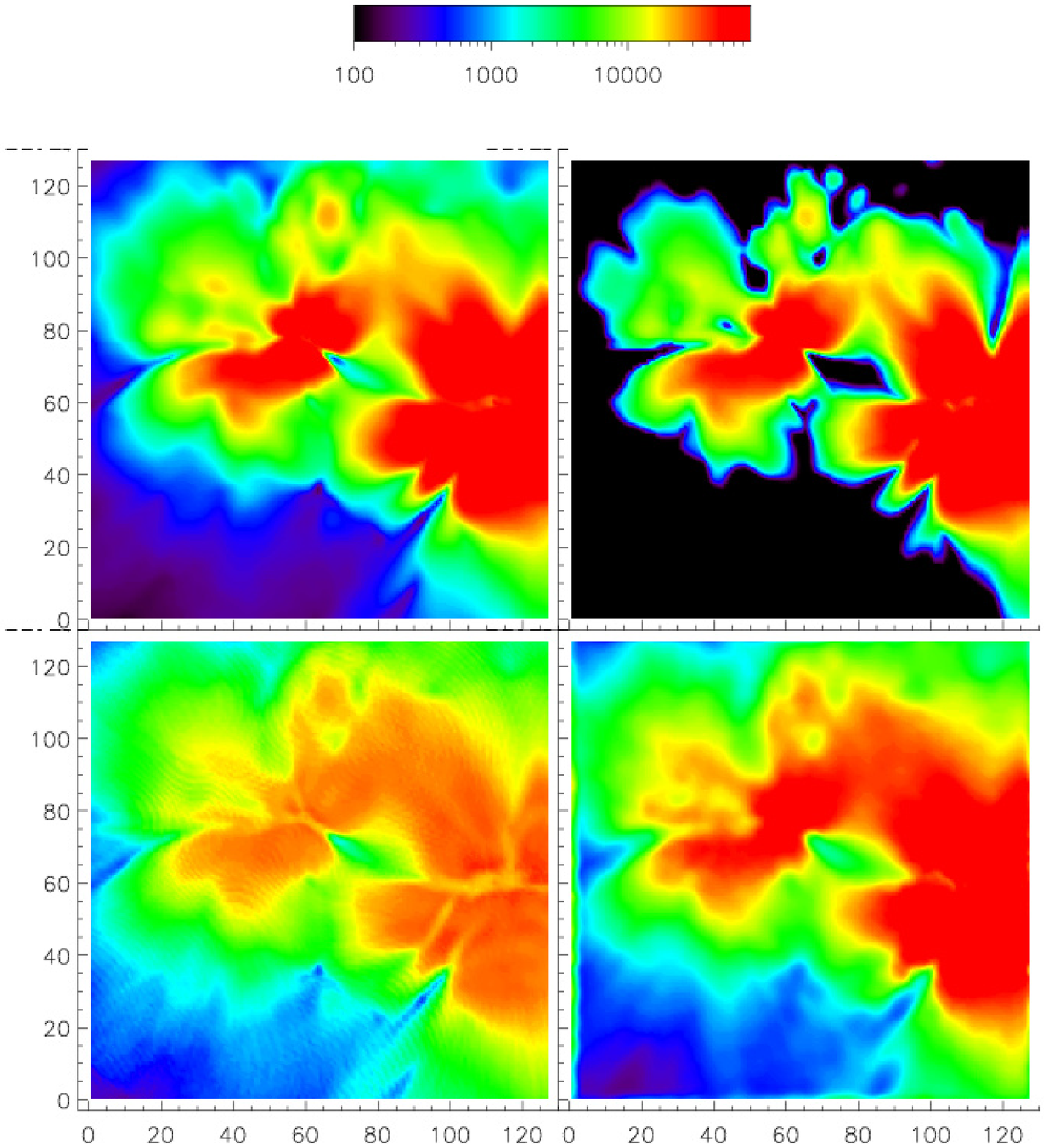,angle=0.,width=9.0truecm}
\end{center}
\caption{\testfour  Surface cut of the temperature through the middle of the simulation volume at t = 0.05 Myr.  The number of rays traced  $N_{\rm t}$ by \code is $q \times 10^8$ where $q = t/0.4$ is the fraction of elapsed time. 
Beginning in the top left corner and proceeding clockwise are the results from {\small C$^2$-Ray}, {\small FTTE}, \code, and {\small CRASH}.}
\label{T4T1surf}
\end{figure}
\begin{figure}
\begin{center}
\psfig{file=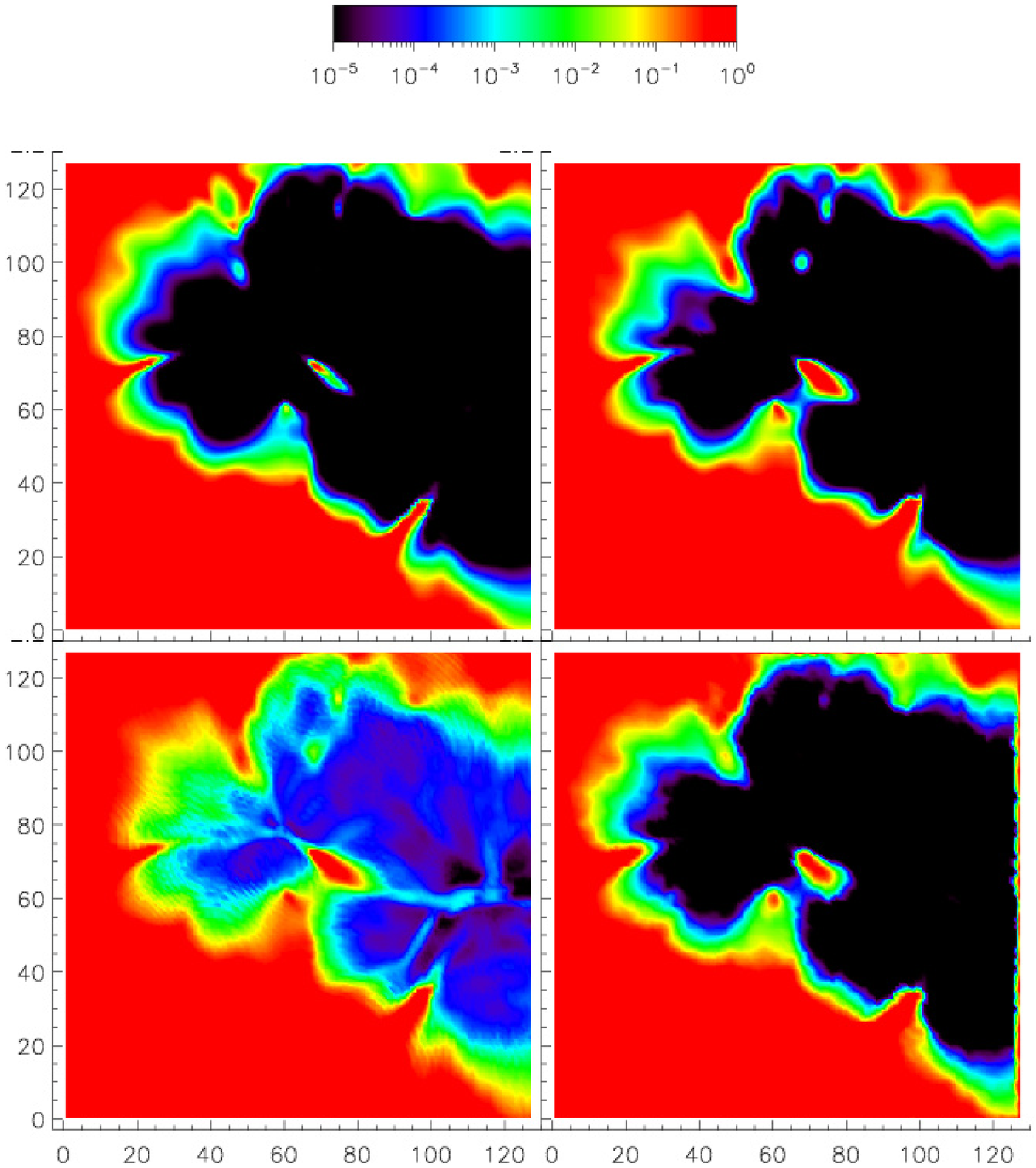,angle=0.,width=9.0truecm}
\end{center}
\caption{\testfour  Surface cut of the neutral fraction through the middle of the simulation volume at t = 0.2 Myr.  The number of rays traced  $N_{\rm t}$ by \code is $q \times 10^8$ where $q = t/0.4$ is the fraction of elapsed time. 
Beginning in the top left corner and proceeding clockwise are the results from {\small C$^2$-Ray}, {\small FTTE}, \code, and {\small CRASH}.}
\label{T4x3surf}
\end{figure}
\begin{figure}
\begin{center}
\psfig{file=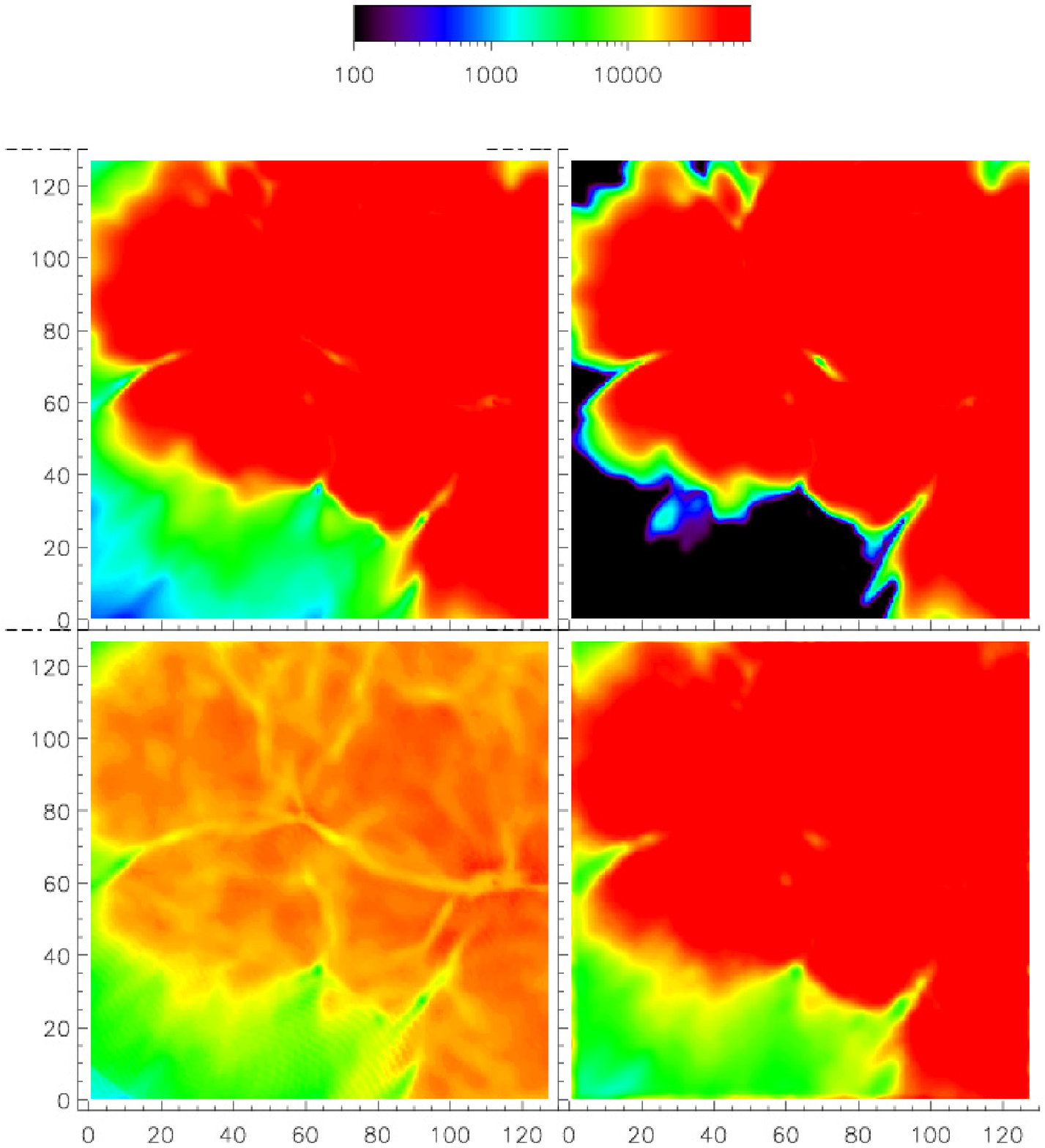,angle=0.,width=9.0truecm}
\end{center}
\caption{\testfour  Surface cut of the temperature through the middle of the simulation volume at t = 0.2 Myr.  The number of rays traced  $N_{\rm t}$ by \code is $q \times 10^8$ where $q = t/0.4$ is the fraction of elapsed time.  
Beginning in the top left corner and proceeding clockwise are the results from {\small C$^2$-Ray}, {\small FTTE}, \code, and {\small CRASH}.}
\label{T4T3surf}
\end{figure}

The results from \code are compared with those of {\small C$^2$-Ray}
\citep{2006NewA...11..374M}, {\small FTTE} \citep{2005MNRAS.362.1413R}, and
{\small CRASH} \citep{ 2003MNRAS.345..379M} in figures \ref{T4x1surf} to
\ref{T4T3surf}.  General features of the ionization field including the extent
and shape of the ionization front, the shadows from dense clumps, and the
shape of the neutral island near the center of the slice, are similar in all
codes.  Although \code and {\small CRASH} share the most similarities,
including the sampling of high energy photons, we see some slight differences
in the peak ionization and temperature values they produce in the central part
of the highly ionized region.

\section{Discussion}

We have presented the cosmological SPH raytracing code \code and discussed the
results of several radiative transfer problems by way of validation. \code
employs a Monte Carlo approach to ray tracing applied directly to the SPH
particle distribution native to a large fraction of current astrophysical and
cosmological hydrodynamic simulations. The column density sums at the heart of
the radiative transfer calculation are carried out
using the SPH kernels so that no
regridding of the data is necessary, maintaining the adaptive Lagrangian nature
that makes SPH attractive in the first place. The statistical nature of the
Monte Carlo method makes the inclusion of arbitrary source spectra and
emission profiles very straightforward.  The simplicity of its implementation
also allows the future addition of more complicated physics as well as
parallelization.  However, a large number of rays must be traced
to get a fair representation of the underlying probability distribution
functions being sampled and to maintain angular resolution.  This translates
to the numerical problem of finding the intersection of numerous rays and
spheres as quickly as possible. In order to do this we have applied a variant
of the neighbor search techniques using oct-trees and a fast box-ray
intersection test adapted from computer graphics, resulting in an efficient
adaptive ray tracing code applicable to current astronomical hydro
simulations.

There are, in general, no analytic solutions to the types of radiative
transfer problems that occur with sources embedded in 3D density fields.
Therefore we validated \code using the tests chosen by the
Radiative Transfer Comparison Project. There is good agreement of \code
results with codes that treat the same level of physics.

We present the source code for \code on a companion 
website\footnote{http://www.sphray.org},
together with a users guide and some example input snapshot files. An early version of this RT approach was used to study the structure of neutral hydrogen
in the Universe at the time of reionization in
\cite{2007arXiv0709.2362C}.

\section*{Acknowledgments}

This project is supported by the National Science Foundation, NSF AST-0205978 and by NASA ATP grant NNG 06-GH88G.  We thank Intel for their 
generous donation of processors used in this work.

\appendix

\section{Photo-ionization cross-sections and Atomic Cooling/Heating Rates}

The rates below make use of the following notation,

\begin{equation}
\lambda_A = 2 \frac{T_A}{T} \nonumber
\end{equation}
 
where the $T_A$ are the ionization energies of the photo absorbing species in temperature units, 

\begin{equation}
T_{\rm HI} = 157,809 {\rm K} \nonumber
\end{equation}

\begin{equation}
T_{\rm HeI} = 285,335 {\rm K} \nonumber
\end{equation}

\begin{equation}
T_{\rm HeII} = 631,515 {\rm K} \nonumber
\end{equation}
and, 

\begin{equation}
T_i = \frac{T}{10^i {\rm K}}
\end{equation}

\begin{itemize}

\item Recombination Rates - Case A \citep{HuiGnedinRates1997} [${ \rm cm^3 \, s^{-1} }$]

\begin{equation}
\alpha^A_{\rm HII} = 1.269 \times 10^{-13} 
\frac{\lambda_{\rm HI}^{1.503}}
{\left[ {1.0 + \left( \frac{\lambda_{\rm HI}}{0.522} \right)^{0.470}} \right]^{1.923}}
\end{equation}

\begin{equation}
\alpha^A_{\rm HeII} = 3.0 \times 10^{-14} \lambda_{\rm HeI}^{0.654} 
\end{equation}

\begin{equation}
\alpha^A_{\rm HeIII} = 2.538 \times 10^{-13} 
\frac{\lambda_{\rm HeII}^{1.503}}
{\left[ {1.0 + \left( \frac{\lambda_{\rm HeII}}{0.522} \right)^{0.470}} \right]^{1.923}}
\end{equation}

\item Recombination Rates - Case B \citep{HuiGnedinRates1997} [${ \rm cm^3 \, s^{-1} }$]

\begin{equation}
\alpha^B_{\rm HII} = 2.753 \times 10^{-14} 
\frac{\lambda_{\rm HI}^{1.500}}
{\left[ {1.0 + \left( \frac{\lambda_{\rm HI}}{2.740} \right)^{0.407}} \right]^{2.242}}
\end{equation}

\begin{equation}
\alpha^B_{\rm HeII} = 1.26 \times 10^{-14} \lambda_{\rm HeI}^{0.750} 
\end{equation}

\begin{equation}
\alpha^B_{\rm HeIII} = 5.506 \times 10^{-14} 
\frac{\lambda_{\rm HeII}^{1.500}}
{\left[ {1.0 + \left( \frac{\lambda_{\rm HeII}}{2.740} \right)^{0.407}} \right]^{2.242}}
\end{equation}

\item Collisional Ionization Rates \citep{CenAtomicRates1992}[${\rm cm^3 \, s^{-1} }$]

\begin{equation}
\gamma_{\rm HI} = \frac{ 5.85 \times 10^{-11} \sqrt{T_0} }
{1 + \sqrt{T_5}} e^{-T_{\rm HI}/T}
\end{equation}

\begin{equation}
\gamma_{\rm HeI} = \frac{ 2.38 \times 10^{-11} \sqrt{T_0} }
{1 + \sqrt{T_5}} e^{-T_{\rm HeI}/T}
\end{equation}

\begin{equation}
\gamma_{\rm HeII} = \frac{ 5.68 \times 10^{-12} \sqrt{T_0} }
{1 + \sqrt{T_5}} e^{-T_{\rm HeII}/T}
\end{equation}

\item Collisional Ionization Cooling \citep{CenAtomicRates1992} [${\rm ergs \, cm^{-3} \, s^{-1}}$]

\begin{equation}
\zeta_{\rm HI} = \frac{ 1.27 \times 10^{-21} \sqrt{T_0} }
{1 + \sqrt{T_5}} e^{-T_{\rm HI}/T} \nel \nHI
\end{equation}

\begin{equation}
\zeta_{\rm HeI} = \frac{ 9.38 \times 10^{-22} \sqrt{T_0} }
{1 + \sqrt{T_5}} e^{-T_{\rm HeI}/T} \nel \nHeI 
\end{equation}

\begin{equation}
\zeta_{\rm HeII} = \frac{ 4.95 \times 10^{-22} \sqrt{T_0} }
{1 + \sqrt{T_5}} e^{-T_{\rm HeII}/T} \nel \nHeII
\end{equation}

\item Collisional Excitation Cooling \citep{CenAtomicRates1992}[${\rm ergs \, cm^{-3} \, s^{-1}}$]

\begin{equation}
\psi_{\rm HI} = \frac{ 7.5 \times 10^{-19} }
{1 + \sqrt{T_5}} e^{-118348/T_0} \nel \nHI
\end{equation}

\begin{equation}
\psi_{\rm HeI} = \frac{ 9.10 \times 10^{-27} T_0^{-0.1687} }
{1 + \sqrt{T_5}} e^{-13179/T_0} \nel^2 \nHeII
\end{equation}

\begin{equation}
\psi_{\rm HeII} = \frac{ 5.54 \times 10^{-17} T_0^{-0.397} }
{1 + \sqrt{T_5}} e^{-473638/T_0} \nel \nHeII
\end{equation}

\item Recombination Cooling - Case A \citep{HuiGnedinRates1997} [${\rm ergs \, cm^{-3} \, s^{-1}}$]

\begin{equation}
\eta^A_{\rm HII} = 1.778 \times 10^{-29} 
\frac{T_0 \lambda_{\rm HI}^{1.965}}
{\left[ {1.0 + \left( \frac{\lambda_{\rm HI}}{0.541} \right)^{0.502}} \right]^{2.697}}
\nel \nHII
\end{equation}

\begin{equation}
\eta^A_{\rm HeII} = k_b T_0 \alpha^A_{\rm HeII} \nel \nHeII
\end{equation}

\begin{equation}
\eta^A_{\rm HeIII} = 1.4224 \times 10^{-28} 
\frac{T_0 \lambda_{\rm HeII}^{1.965}}
{\left[ {1.0 + \left( \frac{\lambda_{\rm HeII}}{0.522} \right)^{0.470}} \right]^{1.923}}
\nel \nHeIII
\end{equation}

\item Recombination Cooling - Case B \citep{HuiGnedinRates1997} [${\rm ergs \, cm^{-3} \, s^{-1}}$]

\begin{equation}
\eta^B_{\rm HII} = 3.435 \times 10^{-30} 
\frac{T_0 \lambda_{\rm HI}^{1.970}}
{\left[ {1.0 + \left( \frac{\lambda_{\rm HI}}{2.250} \right)^{0.376}} \right]^{3.720}}
\nel \nHII
\end{equation}

\begin{equation}
\eta^B_{\rm HeII} = k_b T_0 \alpha^B_{\rm HeII} \nel \nHeII
\end{equation}

\begin{equation}
\eta^B_{\rm HeIII} = 2.748 \times 10^{-29} 
\frac{T_0 \lambda_{\rm HeII}^{1.970}}
{\left[ {1.0 + \left( \frac{\lambda_{\rm HeII}}{2.250} \right)^{0.376}} \right]^{3.720}}
\end{equation}

\item Bremsstrahlung Cooling \citep{CenAtomicRates1992} [${\rm ergs \, cm^{-3} \, s^{-1}}$]

\begin{equation}
\beta = 1.42 \times 10^{-27} g_{ff} \sqrt{T_0} (\nHII + \nHeII + 4\nHeIII) \nel
\end{equation}

where $g_{ff} = 1.5$ is the Gaunt factor.

\item Compton Heating/Cooling \citep{HaimanRates1996} [${\rm ergs \, cm^{-3} \, s^{-1}}$]

\begin{equation}
\chi = 1.017 \times 10^{-37}  T_{\gamma}^4 
\left( T_0 -  T_{\gamma} \right)  \nel 
\end{equation}

where $T_{\gamma} = T_{Bkgnd} / {\rm K}$ is the unitless background radiation temperature.

\end{itemize}

\label{lastpage}

\bibliographystyle{mnras}	
\bibliography{astrobibl}	

\end{document}